\documentclass[reprint,pre,aps,superscriptaddress]{revtex4-1}
\usepackage[utf8]{inputenc}
\usepackage{graphicx}
\usepackage{amsmath}
\usepackage{bbm}
\usepackage{amssymb}
\usepackage{amsfonts}
\usepackage[colorlinks,linkcolor=red,urlcolor=magenta,citecolor=blue]{hyperref}

\newcommand{\Tr}{\mathrm{Tr}}
\newcommand{\ket}[1]{{\left \lvert#1\right\rangle}}
\newcommand{\bra}[1]{{\left\langle#1\right\rvert}}

\begin{document}

\title{Quantum heat statistics with time-evolving matrix product operators}
\author{Maria Popovic}
\affiliation{School of Physics, Trinity College Dublin, College Green, Dublin 2, Ireland}
\author{Mark T. Mitchison}
\affiliation{School of Physics, Trinity College Dublin, College Green, Dublin 2, Ireland}
\author{Aidan Strathearn}
\affiliation{School of Mathematics and Physics, The University of Queensland, St Lucia, Queensland 4072, Australia}
\author{Brendon W. Lovett}
\affiliation{SUPA, School of Physics and Astronomy, University of St Andrews, St Andrews KY16 9SS, United Kingdom}
\author{John Goold}
\affiliation{School of Physics, Trinity College Dublin, College Green, Dublin 2, Ireland}
\author{Paul R. Eastham}
\affiliation{School of Physics, Trinity College Dublin, College Green, Dublin 2, Ireland}

\begin{abstract}
We present a numerically exact method to compute the full counting statistics of heat transfer in non-Markovian open quantum systems, which is based on the time-evolving matrix product operator (TEMPO) algorithm. This approach is applied to the paradigmatic spin-boson model in order to calculate the mean and fluctuations of the heat transferred to the environment during thermal equilibration. We show that system-reservoir correlations make a significant contribution to the heat statistics at low temperature and present a variational theory that quantitatively explains our numerical results. We also demonstrate a fluctuation-dissipation relation connecting the mean and variance of the heat distribution at high temperature. Our results reveal that system-bath interactions make a significant contribution to heat transfer even when the dynamics of the open system is effectively Markovian. The method presented here provides a flexible and general tool to predict the fluctuations of heat transfer in open quantum systems in non-perturbative regimes.
\end{abstract}

\maketitle

\section{Introduction}

The importance of heat management at the nanoscale has grown in tandem with advances in the fabrication and control of small devices, motivating increasing interest in the non-equilibrium thermodynamics of open quantum systems~\cite{goold,Benenti2017,Thermo_Book2018,Mitchison2019}. For example, quantum thermal machines have been studied in such diverse experimental platforms as single-electron transistors~\cite{Koski2014,Koski2015,Josefsson2018}, trapped ions~\cite{Rossnagel2016,Maslennikov2019,Lindenfels2019}, superconducting circuits~\cite{Ronzani2018}, and spin ensembles~\cite{Peterson2019,Klatzow2019}. Numerous technologically or biologically important systems are also naturally described as quantum heat engines, including lasers~\cite{Scovil1959}, light-emitting diodes~\cite{Santhanam2012}, and light-harvesting complexes~\cite{Dorfman2013,Creatore2013,Killoran2015,Fruchtman2016}. These minuscule machines all operate far from equilibrium and are significantly affected by quantum and thermal noise. Strong coupling may blur the boundary between system and environment~\cite{Jarzynski2017,Talkner2020}, potentially leading to non-Markovian effects~\cite{Rivas2014,Breuer2016} with interesting thermodynamic consequences~\cite{Pezzutto2016,maniscalco,raja,popovic,Zicari2020}. In addition, the importance of fluctuations at small scales means that the statistical character of thermodynamic quantities such as work and heat cannot be ignored~\cite{Esposito2009,Campisi2011}. These features together give rise to a rich and varied phenomenology with important ramifications for emerging quantum technologies.

A crucial limiting factor for the performance of quantum devices is the transfer of heat to and from their surroundings. A detailed understanding of heat transfer is therefore essential to optimise control protocols while minimising wasteful dissipation~\cite{Murphy2019,Dann2019,Pancotti2020}. More generally, heat flux is a fundamental source of irreversibility and entropy production in open quantum systems~\cite{Esposito2010,Deffner2011}. Entropy production limits the efficiency of heat engines and refrigerators~\cite{Pietzonka2018}, determines the energy cost of information erasure~\cite{Goold2015} and feedback control~\cite{Sagawa2008}, constrains current fluctuations far from equilibrium~\cite{Barato2015,Gingrich2016,Guarnieri2019,Timpanaro2019,Hasegawa2019}, and can be directly measured in well controlled quantum settings~\cite{Brunelli2018,Micadei2019,Harrington2019}. However, modelling heat transfer in strongly coupled systems is a difficult theoretical problem because it requires access to the energetics of the bath. On the contrary, the majority of techniques for describing open quantum systems either neglect the environment's dynamics completely or treat it via an effective or approximate description~\cite{deVega}. An accurate, tractable method to predict the fluctuations of heat transfer in generic open quantum systems is still lacking.

Here, we fill this gap by developing an efficient numerical method to compute heat statistics using the path-integral formulation of dissipative quantum mechanics~\cite{feynman}. Previous research has shown that the probability distributions of heat and work can be formally derived within this framework~\cite{Funo2018,Aurell2018,Aurell_2020}. However, a direct evaluation of the corresponding path integral is only possible for a few exactly solvable models, while numerical approaches based on the quasi-adiabatic path integral (QUAPI) method~\cite{Makri1,Makri2} require careful fine-tuning to avoid error accumulation~\cite{strathearnNJP,Kilgour2019}. We solve this problem by generalising the TEMPO algorithm~\cite{strathearnNature} to calculate the characteristic function of energy changes in the bath, equivalent to the Fourier transform of the heat probability distribution. This algorithm exploits a tensor-network representation of the QUAPI propagator to describe complicated non-Markovian evolutions efficiently~\cite{Gribben2020}. As a result, we obtain a flexible and accurate tool to describe fluctuating heat transfer in generic, strongly coupled open quantum systems, which can be extended to deal with time-dependent Hamiltonians~\cite{Fux2021} or multiple baths~\cite{Kilgour2019}. 

The canonical open quantum system comprises a small, few-state system coupled to a bosonic bath. This general setting is known to be amenable to efficient tensor-network descriptions~\cite{Prior2010,Somoza2019}. For the sake of concreteness, in this work we focus on the paradigmatic spin-boson model, which describes quantum dots~\cite{Nazir2009}, ultracold atomic impurities~\cite{Recati2005} and superconducting circuits~\cite{Magazzu2018}, to name just a few examples. We demonstrate our approach by applying it to the non-equilibrium quantum thermodynamics of this important model. We first verify the accuracy of our method by comparison with the exact solution in the limit of the independent boson model. Then we compute the time-dependent heat transfer and its fluctuations across a range of parameters in the unbiased spin-boson model, including the challenging low-temperature and strong-coupling regimes. We interpret our results using the notion of generalised equilibration in strong-coupling thermodynamics~\cite{Talkner2020}, and develop analytical models that quantitatively explain the mean heat exchange in the high-temperature and low-temperature limits. We also show numerically that the heat distribution obeys a fluctuation-dissipation relation (FDR) in the high-temperature limit, which is similar to the well-known FDR of the work distribution~\cite{Jarzynski1997}. Interestingly, our results show that the system-bath interaction energy makes a considerable contribution to the heat statistics, even in the weak-coupling and high-temperature regime where a Markovian description of the system dynamics alone is accurate. This underlines the need to interpret with great care the standard Markovian description of quantum thermodynamics~\cite{Alicki1979}, which is based on properties of the open system alone.

A brief outline of the paper is as follows. In the next section, we introduce the spin-boson model and define the thermodynamic quantities of interest. Details of our numerical method are provided in Sec.~\ref{sec:method}. We then present results for the independent-boson and spin-boson models in Sec.~\ref{sec:results}, before concluding in Sec.~\ref{sec:conclusions}. Units where $\hbar  = 1= k_B$ are used throughout.

\section{Preliminaries}
\label{sec:prelim}

\subsection{Quantum thermodynamics of relaxation processes}
\label{sec:thermodynamics}
We are interested in the non-equilibrium thermodynamics of an open quantum system coupled to a large heat bath. The Hamiltonian of such a system can be written generically as
\begin{equation}
\hat{H}= \hat{H}_S + \hat{H}_B +\hat{H}_{I},\label{eq:generic_hamiltonian}
\end{equation}
where $ \hat{H}_S $ is the free Hamiltonian of the quantum system, $ \hat{H}_B $ is the free Hamiltonian of the environment, and $\hat{H}_{I}$ is the Hamiltonian that describes the interaction between these two components. Following the standard approach~\cite{Leggett1987}, the bath is modelled by an infinite collection of harmonic oscillators coupled linearly to the system, so that
\begin{align}
    \hat{H}_B &= \sum_j \omega_j \hat{a}_j^\dagger \hat{a}_j,\\
    \hat{H}_I & = \hat{S} \otimes \sum_j g_j \left(\hat{a}_j +  \hat{a}_j^\dagger\right).
\end{align}
Here, $\hat{a}_j$ is the annihilation operator for mode $j$ of the bath, $\omega_j$ is the corresponding mode frequency, $g_j$ is a coupling constant, and $\hat{S}$ is an arbitrary system operator. The bath is characterized by its spectral density function \cite{Makri1}
\begin{equation}
J\left(\omega\right)=\sum_{j}g_{j}^{2}\delta\left(\omega-\omega_{j}\right)\label{eq:spec_dens_fun}.
\end{equation}
    
Non-equilibrium processes at the nanoscale feature significant and measurable fluctuations. Therefore, thermodynamic process variables such as work, $W$, and heat, $Q$, must be promoted to stochastic quantities described by the corresponding probability distributions, $P(W)$ and $P(Q)$. Thermodynamic work is associated with changes in the external conditions defining the Hamiltonian, while heat is defined here to be the change in energy of the bath. Operationally, each of these quantities can be extracted from a two-point measurement of $\hat{H}$ (work) or $ \hat{H}_B $ (heat) at the beginning and end of the evolution, either with direct projective measurements~\cite{Talkner2007} or via ancillary probes~\cite{Mazzola2013,Dorner2013,Roncaglia2014,Goold2014}. Therefore, under strong-coupling conditions where the commutator $[ \hat{H}_B , \hat{H}_{I} ]$ is non-negligible, work and heat are simultaneously measurable only if the system-bath interaction vanishes at the beginning and end of the evolution~\cite{Talkner2020}. This is the relevant scenario for cyclic thermal machines, for example, and also the one that we assume here.

Following the above reasoning, we consider the relaxation dynamics starting from a product state 
\begin{equation}\label{eq:product_state}
    \hat{\rho}(0) = \hat{\rho}_S(0) \otimes \hat{\rho}_B(0),
\end{equation}
where $\hat{\rho}_S(t) = \Tr_B[\hat{\rho}(t)]$ is the reduced state of the open quantum system and $\hat{\rho}_B(t)$ is the state of the bath with a thermal initial condition $\hat{\rho}_B(0) = e^{-\beta  \hat{H}_B }/\Tr[e^{-\beta  \hat{H}_B }]$ at inverse temperature $\beta = 1/T$. The system evolves in time according to $\hat{\rho}(t) = \hat{U}(t) \hat{\rho}(0) \hat{U}^\dagger(t)$, where $\hat{U}(t) = e^{-i \hat{H} t}$ is the time evolution operator. The energy and entropy change of the system are given respectively by
\begin{align}
\Delta U & =\langle  \hat{H}_S \rangle _{t}-\langle  \hat{H}_S \rangle _{0},\label{eq:internal_energy_variation}\\
\Delta S&=S\left[\hat{\rho}_{S}(t)\right] - S\left[\hat{\rho}_{S}(0)\right],\label{eq:entropy_change}
\end{align}
where we denote time-dependent expectation values by $\langle  \bullet \rangle_{t} \equiv \Tr[ \bullet \hat{\rho}(t)]$ and $S[\hat{\rho}] = -\Tr [\hat{\rho}\ln \hat{\rho}]$ is the von Neumann entropy of the state $\hat{\rho}$. Note that, unless the initial and final states of the system are in thermal equilibrium, neither $\Delta U$ nor $\Delta S$ as defined above necessarily correspond to variations of thermodynamic potentials.

Since the Hamiltonian is time-independent during the relaxation process, all energy transferred during the evolution is in the form of heat exchanged with the bath. 
The mean heat absorbed by the bath is given by
\begin{equation}
\left\langle Q\right\rangle =\langle  \hat{H}_B \rangle_{t}-\langle  \hat{H}_B \rangle_{0}.\label{eq:heat_change}
\end{equation}
The first law of thermodynamics states that
\begin{equation}\label{eq:first_law}
  \langle Q\rangle = \langle W\rangle - \Delta U,
\end{equation}
where $ \langle W\rangle$ is the average work performed on the entire system by switching the system-bath interaction on and off at the endpoints of the evolution. Assuming that this switching is instantaneous, we have that $\langle W\rangle  = - \langle  \hat{H}_{I} \rangle_t$ (i.e., $\langle  \hat{H}_{I} \rangle_t$ is the mean interaction energy just before it is switched off), which follows from energy conservation, $\langle \hat{H}\rangle_t = \langle \hat{H}\rangle_0$, and the fact that $\langle  \hat{H}_{I} \rangle_0= 0$ for an interaction of the form of Eq.~\eqref{eq:interHam}. The average heat dissipated into the bath therefore comprises two contributions: the change in the system's internal energy and the system-bath interaction energy developed throughout the relaxation process. This dissipation is associated with an average entropy production
\begin{equation}
    \label{eq:second_law}
    \langle \Sigma \rangle  = \Delta S + \beta \langle Q\rangle,
\end{equation}
which obeys $\langle \Sigma\rangle \geq 0$ in accordance with the second law~\cite{Esposito2010,Deffner2011}, where equality holds for reversible processes.


\subsection{Heat statistics}
\label{sec:charfun}

By definition, the heat transfer is the energy change that would be registered by projective energy measurements on the bath at the beginning and end of the process. We denote by $\hat{\Pi}_n = \ket{E_n}\bra{E_n}$ the projector onto the eigenstate $\ket{E_n}$ of $\hat{H}_B$ with eigenvalue $E_n$. The heat distribution is then defined by
\begin{equation}
    \label{heat_distro}
    P(Q) = \sum_{m,n} p_n p_{n\to m} \delta (Q+E_n-E_m),
\end{equation}
where $p_n = {\rm Tr}[(\mathbbm{1}\otimes \hat{\Pi}_n) \hat{\rho}(0)]$ is the probability of measuring initial energy $E_n$, and $p_{n\to m} = {\rm Tr} [ \hat{\Pi}_m \hat{U}(t) (\hat{\rho}_S(0)\otimes\hat{\Pi}_n)\hat{U}^\dagger(t)]$ is the conditional probability for the transition $E_n\to E_m$ ~\cite{Funo2018}. The fluctuating heat exchange can be characterised by the statistical moments
\begin{align}
    \langle Q^n\rangle & = \int_{-\infty}^\infty dQ \, P(Q) Q^n \label{eq:heat_moments_P}\\ 
    & = (-i)^n\left.\frac{d^n}{du^n} \chi(u)\right\rvert_{u=0}. \label{eq:heat_moments_chi}
\end{align}
Here, we have introduced the characteristic function 
\begin{equation}
\chi(u)=\int_{-\infty}^{\infty}dQ\,P(Q)e^{iuQ},\label{eq:char_func}
\end{equation}
where $u$ is the counting field parameter. Using Eq.~\eqref{heat_distro}, one easily obtains~\cite{Esposito2009}
\begin{equation}
\chi(u) = \Tr\left[e^{iu \hat{H}_B }\hat{U}(t)e^{-iu \hat{H}_B }\hat{\rho}(0)\hat{U}^\dagger(t)\right].\label{eq:char_fun_2}
\end{equation}
It is convenient to define a modified time evolution operator as
\begin{equation}
\label{eq:modified_TE}
\hat{V}_{u}(t)=e^{i  \hat{H}_B u/2}\hat{U}(t)e^{-i \hat{H}_B u/2}.
\end{equation}
This allows us to rewrite Eq.~\eqref{eq:char_fun_2} as $\chi(u)=\Tr\left[\hat{\rho}(t,u)\right]$, with the modified density matrix
\begin{equation}
\hat{\rho}(t,u)=\hat{V}_{u}(t)\hat{\rho}(0)\hat{V}_{-u}^{\dagger}(t).\label{eq:modified_dm}
\end{equation}
Defining $\hat{\rho}_{S}(t,u)=\Tr_{B}\left[\hat{\rho}(t,u)\right]$ as the reduced modified system density matrix, we have
\begin{equation}
\chi(u)=\Tr_{S}\left[\hat{\rho}_{S}(t,u)\right]\label{eq:char_fun_3}.
\end{equation}
The form in Eq.~(\ref{eq:char_fun_3}) facilitates the calculation of the heat statistics by means of path-integral techniques.
\\

\section{Path Integral Methods}
\label{sec:method}
\subsection{Influence functional for the modified density matrix}

The dynamics of the modified reduced density matrix $\hat{\rho}_{S}(t,u)$ can be formulated as a path integral~\cite{feynman}, in which the effects of the environment on the open quantum system are captured by an influence functional that is non-local in time. This Feynman-Vernon influence functional is well suited to numerically discretised approaches such as QUAPI~\cite{Makri1,Makri2}, upon which the TEMPO algorithm is built~\cite{strathearnNature}. Here we describe how to obtain the influence functional modified by the counting field $u$.

To derive a discretised path integral for the modified density matrix, $\hat{\rho}_S(t,u) = \Tr_B\left[\hat{V}_u(t)\hat{\rho}(0)\hat{V}_{-u}^\dagger(t)\right]$, we divide the time interval of interest $t$ into $N$ intervals of equal length $\Delta$, as $t=N\Delta$. Then the total time evolution operator
can be expressed as $\hat{V}_u(t)=\left(e^{-i\hat{H}_u\Delta }\right)^{N}$, with $\hat{H}_u = e^{i\hat{H}_Bu/2} \hat{H}e^{-i\hat{H}_Bu/2}$ the Hamiltonian dressed by the counting field. The environmental degrees of freedom are separated from those of the system by defining the Hamiltonian $\hat{H}^{\rm env}_u = \hat{H}_u- \hat{H}_{S}$. The evolution operator over each time interval is then approximated by the symmetric Trotter splitting
\begin{equation}
e^{-i\hat{H}_u\Delta} = e^{-i \hat{H}_S \Delta/2}e^{-i \hat{H}^{\rm env}_{u} \Delta}e^{-i \hat{H}_S \Delta/2} + \mathcal{O}\left(\Delta^{3}\right).\label{eq:trotter splitting}
\end{equation}
The path integral for $\hat{\rho}_S(t,u)$ is constructed by inserting resolutions of the identity in the eigenbasis of the system coordinate $\hat{S}$ at each time step and then tracing over the bath. We use the notation $\ket{s_k^\pm}$ for the eigenstates of $\hat{S}$, where the index $k$ indicates the time $t_k = k\Delta$ and the superscript $+$ ($-$) is used to label eigenvectors inserted on the left (right) of the density matrix. Given our product initial condition in Eq.~\eqref{eq:product_state}, we find
\begin{align}
\left\langle s_{N}^{+}\right|\hat{\rho}_{S}(t,u)\left|s_{N}^{-}\right\rangle =\sum_{s_{0}^{\pm},s_{1}^{\pm}...s_{N-1}^{\pm}}&F\left(\left\{ s_{k}^{\pm}\right\} \right)I\left(\left\{ s_{k}^{\pm}\right\},u\right)\notag\\
&\times\bra{s_{0}^{+}}\hat{\rho}'_{S}(0)\ket{s_{0}^{-}}\label{eq:discrete_path_integral}.
\end{align}
Here, $\hat{\rho}_S'(0) = e^{-i\hat{H}_S\Delta/2}\hat{\rho}_S(0)e^{i\hat{H}_S\Delta/2}$ is a modified initial condition, $F\left(\left\{ s_{k}^{\pm}\right\}\right) = \prod_{k=1}^N G(s^\pm_k, s^\pm_{k-1})$ is a product of free propagators for the system, with
\begin{align}\label{free_propagator}
    G(s^\pm_k, s^\pm_{k-1}) = \bra{s^+_k}e^{-i\hat{H}_S\Delta_k}\ket{s^+_{k-1}}\bra{s^-_{k-1}}e^{i\hat{H}_S\Delta_k}\ket{s^-_{k}},
\end{align}
where $\Delta_k = \Delta$ for $k<N$ and $\Delta_N = \Delta/2$, while the modified influence functional is 
\begin{widetext}
\begin{equation}
    \label{influence_functional}
I\!\left(\left\{ s_{k}^{\pm}\right\},u\right) = \Tr_B \left[  \prod_{k=1}^{N} e^{-i \hat{H}^{\rm env}_{u}(s_{N-k}^+)\Delta}\hat{\rho}_B(0) \!\prod_{k'=0}^{N-1} e^{i \hat{H}^{\rm env}_{-u}(s_{k'}^-)\Delta} \right].
\end{equation}
\end{widetext}
Above, we have defined $\hat{H}^{\rm env}_u(s) = \bra{s} \hat{H}_u^{\rm env}\ket{s}$ as the environment Hamiltonian conditioned on a particular eigenvalue $s$ of the system coordinate.

To evaluate the influence functional explicitly for the spin-boson model, we introduce a compact superoperator notation~\cite{strathearnNJP,Aurell_2020}. Let us move to an interaction picture with respect to the free Hamiltonian $\hat{H}_0 = \hat{H}_S + \hat{H}_B$ by writing $\tilde{\rho}(t,u) = e^{i\hat{H}_0t}\hat{\rho}(t,u)e^{-i\hat{H}_0t}$. From Eq.~\eqref{eq:modified_dm}, we derive the differential equation $(d/dt)\tilde{\rho}(t,u) = \mathcal{L}_I(t,u)\tilde{\rho}(t,u)$, where the Liouvillian superoperator is defined by
\begin{equation}\label{modified_Liouvillian}
    i \mathcal{L}_I(t,u)\bullet = \tilde{H}_I(t,u) \bullet \, - \, \bullet \tilde{H}_I(t,-u) ,
\end{equation}
with $\tilde{H}_I(t,u) =  e^{i\hat{H}_Bu/2}\tilde{H}_I(t) e^{-i \hat{H}_Bu/2}$ and $\tilde{H}_I(t) = e^{i\hat{H}_0t}\hat{H}_I e^{-i \hat{H}_0t}$. The solution for the modified reduced density matrix is 
\begin{equation}
    \label{red_mod_dm_soln}
    \tilde{\rho}_S(t,u) = \mathcal{I}(t,u)\hat{\rho}_S(0),
\end{equation}
where the influence superoperator is given by
\begin{equation}
\label{I_superop_def}
    \mathcal{I}(t,u) = \left\langle \overleftarrow{T} \!\exp \left[ \int_0^t dt'\mathcal{L}_I(t',u)\right] \right\rangle_B,
\end{equation}
and we introduced the time-ordering symbol $\overleftarrow{T}$, which reorders superoperators such that time increases from right to left, and the reservoir average, for any superoperator $\mathcal{X}$,
\begin{equation}
    \label{reservoir_average}
    \left\langle \mathcal{X} \right\rangle_B \equiv \Tr_B \left[ \mathcal{X}\hat{\rho}_B(0) \right].
\end{equation}
Since the interaction Hamiltonian $\hat{H}_I$ is linear and the reservoir thermal state is Gaussian, we may express Eq.~\eqref{I_superop_def} exactly using a time-ordered cumulant expansion up to second order~\cite{Kubo1962}:
\begin{equation}
\mathcal{I}(t,u)=\overleftarrow{T}\!\exp\left[\int_{0}^{t}dt'\!\int_{0}^{t'}dt''\left\langle \mathcal{L}_{I}(t',u)\mathcal{L}_{I}(t'',u)\right\rangle _{B}\right].\label{eq:mod_IF}
\end{equation}
The exponent is evaluated using well-known properties of bosonic thermal states; see Appendix~\ref{app:SU_methods} for details. The result is expressed in terms of three correlation functions:
\begin{align}
\eta^{\mathcal{C}}(t,u) & =\int_{0}^{\infty}d\omega\frac{J(\omega)}{2\omega^{2}}\sin(u\omega)\label{eq:cont_etaC}\\
&\times\left[\coth\left(\frac{\omega}{2T}\right)[\sin(\omega t)-\omega t]-i\left[1-\cos(\omega t)\right]\right]\nonumber,
\end{align}
\begin{align}
\eta^{\mathcal{A}_{1}}(t,u) & =\int_{0}^{\infty}d\omega\frac{J(\omega)}{\omega^{2}}\cos^{2}\left(\frac{u\omega}{2}\right)\label{eq:cont_etaA1}\\
&\times\left[\coth\left(\frac{\omega}{2T}\right)\left[1-\cos(\omega t)\right]+i\left[\sin(\omega t)-\omega t\right]\right]\nonumber,
\end{align}
\begin{align}
\eta^{\mathcal{A}_{2}}(t,u) & =\int_{0}^{\infty}d\omega\frac{J(\omega)}{\omega^{2}}\sin^{2}\left(\frac{u\omega}{2}\right)\label{eq:cont_etaA2}\\
&\times\left[\coth\left(\frac{\omega}{2T}\right)\left[1-\cos(\omega t)\right]+i\left[\sin(\omega t)-\omega t\right]\right]\nonumber.
\end{align}
where $J(\omega)$ is the spectral density function of the bath defined in Eq.~\eqref{eq:spec_dens_fun}.
Following Ref.~\cite{strathearnNJP}, we recover the path integral representation from Eq.~\eqref{red_mod_dm_soln} by simply discretising time into $N$ intervals and inserting resolutions of the identity at each time step, $\hat{1} = \sum_{s_k^\pm} \ket{s_k^\pm}\bra{s_k^\pm}$, as in Eq.~\eqref{eq:discrete_path_integral}. In the interaction picture, the free propagators $F(\lbrace s^\pm_k\rbrace)$ do not appear and we obtain the influence functional in the form
\begin{align}
    \label{influence_functional_total}
    I\left(\left\{ s_{k}^{\pm}\right\},u\right) & = \prod_{k=0}^N \prod_{k'=0}^k I_{\Delta k}(s_k^\pm, s_{k'}^\pm,u), \\
    \label{influence_functional_pairs}
    I_{\Delta k}\left(s_k^\pm, s_{k'}^\pm,u\right) & = \exp \left[ - \sum_{q,q' = \pm} s_k^q  \eta^{qq'}_{k-k'}(u)  s^{q'}_{k'}  \right].
\end{align}
Here, $\Delta k = k-k'$ and $\eta^{qq'}_{k-k'}(u)$ are the discretised correlation functions 
\begin{align}
    \label{eta_pp}
    \eta^{++}_{k-k'}(u)&  = \eta^{\mathcal{A}_1}_{k-k'}(u) + \eta^{\mathcal{A}_2}_{k-k'}(u)= \left[\eta^{--}_{k-k'}(u)\right]^* \\
    \eta^{-+}_{k-k'}(u) & = \eta^{\mathcal{A}_2}_{k-k'}(u) - \eta^{\mathcal{A}_1}_{k-k'}(u) +2\eta^{\mathcal{C}}_{k-k'}(u)  \\  
    \eta^{+-}_{k-k'}(u) & =\left[  \eta^{\mathcal{A}_2}_{k-k'}(u) - \eta^{\mathcal{A}_1}_{k-k'}(u) -2\eta^{\mathcal{C}}_{k-k'}(u) \right]^*
\end{align}
where $\eta^{\alpha}_{k-k'}(u) = \eta^\alpha(t_k - t_{k'},u)$, for $\alpha = \mathcal{C}, \mathcal{A}_1, \mathcal{A}_2$. Our expression for $I\left(\left\{ s_{k}^{\pm}\right\},u\right)$ matches the one recently derived in Ref. \cite{Kilgour2019} and it is straightforward to verify that, for $u=0$, it reduces to the original influence functional described in Ref.~\cite{Makri1}. 

The form of Eq.~\eqref{influence_functional_total} emphasises that the environment introduces memory into the evolution by coupling the system coordinate to itself at different times. Crucially, however, the correlation functions $\eta^\alpha(t,u)$ decay to zero for sufficiently large $t$ and therefore the memory time of the environment is finite. This insight forms the basis of the TEMPO algorithm described in the following section. 

\subsection{TEMPO algorithm}
\label{sec:TEMPOalgorithm}

TEMPO~\cite{strathearnNature} is an efficient algorithm to compute path sums of the form of Eq.~\eqref{eq:discrete_path_integral}, given an influence functional of the form of Eq.~\eqref{influence_functional_total}. The standard TEMPO algorithm can be applied directly to our problem, with the only novelty being that here the influence functional is parametrised by the counting field $u$. We therefore provide only a brief summary of TEMPO here, directing the interested reader to Ref.~\cite{strathearnNature} for a detailed description.

The key assumption of both the QUAPI and TEMPO methods is that the non-local time correlations encoded in the influence functional have a finite range, i.e.\ $\eta^\alpha(t,u) \approx 0$ for $t>\tau_C$, where $\tau_{C}$ is the bath memory time. Therefore, in the discretised form of the modified influence functional~\eqref{influence_functional_total} one can introduce a maximum value of $\left|k-k'\right|$ beyond which the coefficients $\eta^\alpha_{k-k'}(u)$ are negligible for all $u$. As a result, we may approximate $I_{\Delta k}(s_{k}^{\pm},s_{k'}^{\pm},u) \approx 1$ for $|k-k'| > K$, where the memory depth $K$ is chosen to be at least $K \geq \tau_{C}/\Delta$. 

The assumption of finite memory depth allows for an efficient description of the quantum dynamics through an iterative tensor propagation scheme, which forms the basis of QUAPI~\cite{Makri1}. To see this, note that the summand in Eq.~\eqref{eq:discrete_path_integral} can be viewed as an $(N+1)$-index object called the augmented density tensor (ADT), denoted $A^{\sigma_N \cdots \sigma_1 \sigma_0}$, where each ``super-index'' $\sigma_k = \{s^+_k,s_k^-\}$ takes four possible values (there are $d^2$ values in general, with $d$ the dimension of the system $S$). The modified density matrix is found by summing over all but the final index, i.e. 
\begin{equation}
    \bra{s_N^+}\hat{\rho}_S(t,u)\ket{s^-_N} = \sum_{\sigma_0,\cdots, \sigma_{N-1}} A^{\sigma_N\cdots \sigma_0},
\end{equation}
where the remaining index $\sigma_N = \{s_N^+,s_N^-\}$ is determined by the values of $s_N^\pm$ on the left-hand side. The ADT is built iteratively starting from the initial condition $A^{\sigma_0} = I_0(s^\pm_0,u) \bra{s^+_0}\hat{\rho}_S(0)\ket{s^-_0}$. Defining the propagator tensors
\begin{align}
\label{propagator_tensor}
    B^{\sigma_n\cdots \sigma_0}_{\mu_{n-1}\cdots \mu_0} & = \left(\prod_{k=1}^{n} \delta^{\sigma_{n-k}}_{\mu_{n-k}}\right)G(s^\pm_n,s^\pm_{n-1}) \notag \\ & \qquad \times \prod_{\Delta k = 0}^n I_{\Delta k}(s^\pm_n, s^\pm_{n-\Delta k},u),
\end{align}
with $\delta^{\sigma}_{\mu}$ the Kronecker delta symbol, the ADT at the $n$th time step is given by the contraction
\begin{equation}
 \label{tensor_iteration}
 A^{\sigma_n \cdots\sigma_0} = B^{\sigma_n\cdots \sigma_0}_{\mu_{n-1}\cdots\mu_0}A^{\mu_{n-1} \cdots\mu_0},
\end{equation}
with the Einstein summation convention assumed. Due to the finite memory depth $K$, the propagator~\eqref{propagator_tensor} acts non-trivially on at most $K$ indices of the ADT, since $I_{\Delta k}(s^\pm_n, s^\pm_{n-\Delta k}) = 1$ for $\Delta k>K$. At the $n$th time step, therefore, when $n>K$ one needs only to store the object $A^{\sigma_{n}\cdots\sigma_{n-K}}$, with the remaining indices summed over. (For the first $K$ time steps one stores the full ADT.)

The limiting factor for QUAPI is the computational resources needed to store and perform contractions on $K$-index tensors. The TEMPO approach circumvents this limitation by representing the ADT and the propagators as tensor networks, which can be stored efficiently using truncated singular-value decompositions, enabling very large values of $K$ to be reached. Comparing with previously developed methods, which are able to perform at values up to $K\sim10$ ~\cite{Kilgour2019}, in the calculations we show in this paper the TEMPO approach performs at a value of $K=500$. The tensor-network representation is efficient due to the finite range of temporal correlations contained in the ADT. This is analogous to the well-known ability of tensor networks to represent many-body quantum states exhibiting short-ranged spatial correlations~\cite{Schollwoeck2011}. In the present case, the bond dimension, i.e.~the number of singular values retained during the construction of the tensor network, quantifies correlations between different time points induced by the non-Markovian environment. The bond dimension is controlled by retaining only those singular values $\lambda$ greater than a cutoff $\lambda_C$.  We define the cutoff as $\lambda_{C}=\lambda_{\rm max}10^{-p/10}$, with $\lambda_{\rm max}$
the highest singular value. The accuracy of the algorithm is therefore controlled by the  exponent $p$ as well as the memory depth $K$ and the numerical time step $\Delta$. 

\section{Spin-Boson Model Results}
\label{sec:results}

Although our method is general, in the following we specialise to the spin-boson model describing a single spin one-half interacting with a bosonic bath of harmonic oscillators~\cite{Leggett1987}. In this case, the terms in Eq.~\eqref{eq:generic_hamiltonian} take the form
\begin{align}
 \hat{H}_S &=\omega_{0}\hat{S}_z+\Omega \hat{S}_x,\label{eq:systemHam}\\
 \hat{H}_B &=\sum_{j}\omega_{j} \hat{a}_{j}^{\dagger}\hat{a}_{j},\label{eq:bathHam}\\
\hat{H}_{I}&=\hat{S}_z\sum_{j}g_{j}\left(\hat{a}_{j}+\hat{a}_{j}^{\dagger}\right).\label{eq:interHam}
\end{align}
Above, $\hat{S}_z$ and $\hat{S}_x$ are the spin operators for the system. We focus on an Ohmic spectral density function of the form 
\begin{equation}
\label{eq:ohmic spectral density}
J\left(\omega\right)=2\alpha\omega e^{-\omega/\omega_{C}},
\end{equation}
where $\alpha$ is a dimensionless coupling constant and $\omega_{C}$ is a large cutoff frequency. In the following, we consider two different limits of the spin-boson model: the independent boson model with $\Omega = 0$, and the unbiased spin-boson model with $\omega_0 = 0$ and $\Omega\neq 0$. The independent boson model is exactly solvable, allowing us to verify the accuracy of our numerical method. We then turn to the unbiased spin-boson model, an archetypal example of a non-integrable open quantum system.

\subsection{Independent boson model}
\label{sec:independent_boson}

The independent boson (IB) model is described by Eqs.~\eqref{eq:systemHam}--\eqref{eq:interHam} with $\Omega=0$. The Hamiltonian can be diagonalised by a polaron transformation, which takes the general form
\begin{equation}
    \label{polaron_transformation}
    \hat{P} = \exp \left[\hat{S}_z\sum_j \frac{f_j}{\omega_j} \left(\hat{a}_j - \hat{a}_j^\dagger\right)\right].
\end{equation}
This describes a spin-dependent displacement of each bath oscillator by an amount proportional to $f_j$. The choice $f_j = g_j$ diagonalises the IB Hamiltonian as $\hat{P}^\dagger \hat{H} \hat{P} = \hat{H}_0 - \tfrac{1}{2}E_r$, where $\hat{H}_0 =  \hat{H}_S  +  \hat{H}_B $ is the free Hamiltonian and we have defined the reorganisation energy
\begin{equation}
    E_r = \frac{1}{2}\int_0^\infty d\omega\, \frac{J(\omega)}{\omega} = \alpha \omega_C,
\end{equation}
which determines the shift in ground-state energy due to the system-bath interaction.

\begin{figure}
\begin{centering}
\includegraphics[scale=0.5]{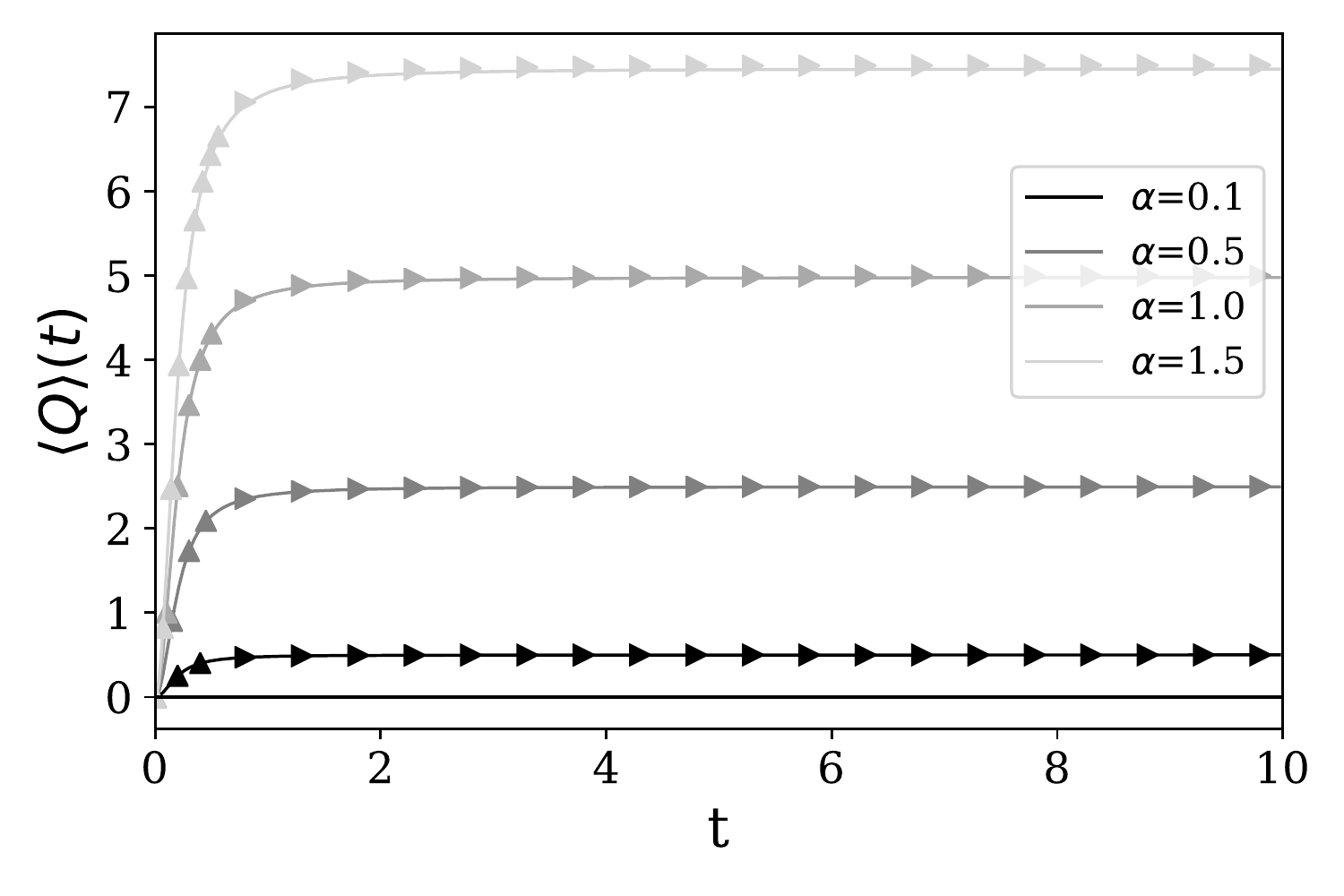}
\par\end{centering}
\caption{\label{fig:independentboson} Mean heat dissipated into the bath as a function of time in the independent boson model, as given by  Eq.~\eqref{mean_heat_IBM} (triangles) and as calculated numerically (solid lines), for four different values of the coupling strength $\alpha$. The spin splitting is $\omega_0 = 1$, the temperature is $T=5$, and the bath cutoff is $\omega_{C}=5$. The parameters controlling the numerical accuracy are $K\Delta=5$, $\Delta=0.01$, $p=100$, and the derivative is taken at $u=0.01$.}
\end{figure}

In the IB model, $[\hat{H}, \hat{H}_S ]=0$, meaning that the local energy of the spin is conserved and $\Delta U = 0$. Therefore, the heat dissipated into the bath is associated purely with the system-bath interaction, as discussed in Sec.~\ref{sec:thermodynamics}. In particular, we show in Appendix~\ref{app:chi_IBM} that the heat characteristic function is independent of the state of the spin and given explicitly by 
\begin{align}
    \label{chi_IBM}
    \ln\chi(u) & = -\frac{1}{2}\int_0^\infty d\omega\, \frac{J(\omega)}{\omega^2}\left[1-\cos(\omega t)\right] \\ 
    & \qquad \times \left\lbrace\left[1-\cos(\omega u)\right]\coth\left(\frac{\omega}{2T}\right) - i \sin(\omega u)\right\rbrace.\notag
\end{align}
Differentiation of this quantity yields closed-form expressions for arbitrary cumulants of the heat distribution, given in Appendix~\ref{app:chi_IBM}. Specifically, the mean heat is found to be
\begin{align}
\label{mean_heat_IBM}
    \langle Q\rangle & = \frac{1}{2}\int_0^\infty d\omega \frac{J(\omega)}{\omega}\left[1-\cos(\omega t)\right], 
\end{align}
which is strictly positive and independent of temperature. Interestingly, these properties are shared by all odd cumulants of the heat distribution in the IB model. For an Ohmic spectral density, we have $\langle Q\rangle = \alpha \omega_C^3 t^2/(1+\omega_C^2t^2)$, which monotonically approaches the reorganisation energy in the long-time limit:
\begin{equation}
\label{asymptotic_heat_IBM}
    \langle Q\rangle_\infty = \alpha\omega_C = E_r.
\end{equation}

For an Ohmic spectral density function, Eq.~\eqref{mean_heat_IBM} depends on only two parameters, the coupling strength and the frequency cutoff. While $\omega_C$ sets the timescale of the heat transfer process, the mean exchanged heat scales linearly with $\alpha$. At first glance, it is not obvious that for strong coupling our method will be able to give the correct prediction, as this regime is in general difficult to model. It is therefore of interest to demonstrate the validity of the numerical method for different values of $\alpha$.

The mean heat is plotted as a function of evolution time for several different coupling strengths in Fig.~\ref{fig:independentboson}. We use these results to validate the numerical algorithm, whose results are shown in the same plot. We find excellent agreement between our simulations and the exact solution for each value of $\alpha$ considered. A simple estimate of the accuracy of our approach is obtained by comparing the asymptotic heat values to the exact result in Eq.~\eqref{asymptotic_heat_IBM}. For the convergence parameters we have used, we find a relative discrepancy of $\delta Q/Q = 0.04\%$ in the case of $\alpha=0.1$, which increases to $\delta Q/Q=0.67\%$ in the case of $\alpha=1.5$. These discrepancies could be further reduced by increasing the accuracy of TEMPO through changing the convergence parameters $\Delta$, $p$ and $K$. For an in depth discussion on the accuracy of the mean heat calculations with respect to the convergence parameters and value of the counting field, see Appendix~\ref{app:convergence_parameters}.

\begin{figure}
\begin{centering}
\includegraphics[scale=0.5]{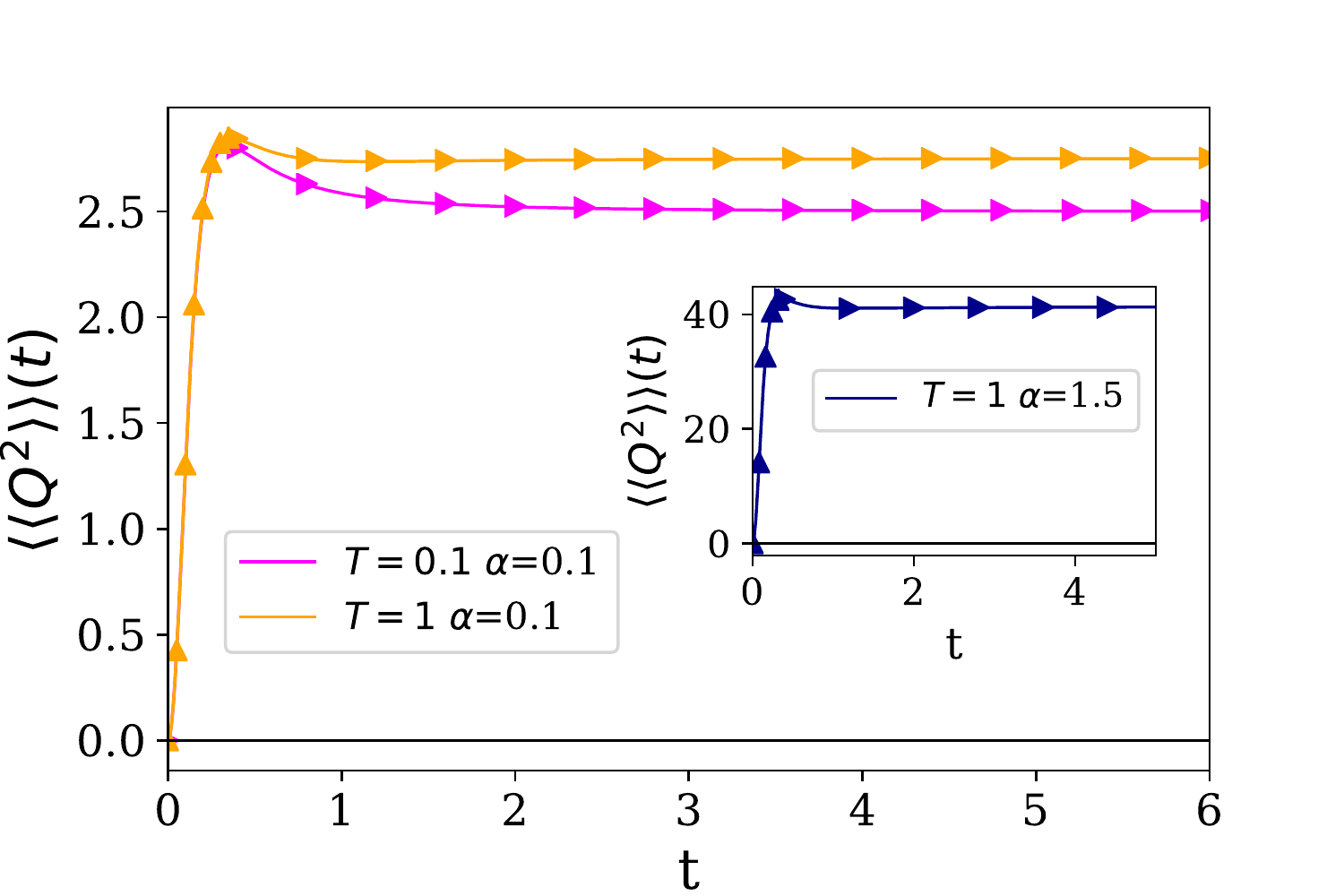}
\par\end{centering}
\caption{\label{fig:secondcumulantIBM} Variance of the heat dissipated into the bath as a function of time in the independent boson model. The solid lines are the second cumulant calculated numerically for the values of temperature and coupling strength indicated. The triangular markers are the corresponding analytical results given by Eq.~\eqref{second_cumulant_IBM}, $l=1$. The spin splitting is set to $\omega_0 = 1$ and the bath cutoff is $\omega_{C}=5$. The parameters controlling the numerical accuracy are $K\Delta=5$, $\Delta=0.01$, and $p=100$. The derivative is taken at $u=0.01$ for $\alpha=0.1$ and at $u=0.005$ for $\alpha=1.5$.}
\end{figure}

To quantify the fluctuations of the exchanged heat, we consider the variance $\langle\!\langle Q^2\rangle\!\rangle = \langle Q^2\rangle -\langle Q\rangle^2$, which is given by
\begin{align}
\label{variance_IBM}
\langle\!\langle Q^2\rangle\!\rangle & = \frac{1}{2}\int_0^\infty d\omega J(\omega) \left[1-\cos(\omega t)\right]\coth\left (\frac{\beta\omega}{2}\right).
\end{align}
Unlike the mean heat in Eq.~\eqref{mean_heat_IBM}, which is independent of temperature, the variance in Eq.~\eqref{variance_IBM} depends on the inverse temperature of the bath $\beta$. We show that our method is accurate for both a lower and a comparable temperature $k_{B}T$ with respect to the energy scale of the system $\omega_0$. Fig.~\ref{fig:secondcumulantIBM} shows the variance as a function of time, for different values of temperature and coupling strength. The numerical predictions again match the analytical solutions given by Eq.~\eqref{variance_IBM}. Note that in order to get a better match between the solutions for high coupling, $\alpha=1.5$, the value of the counting field at which the numerical derivative of $\chi(u)$ is taken has been set to $u=0.005$, compared to the value of $u=0.01$ in the case of $\alpha=0.1$. This suggests that high coupling strength cases require in general more computational precision than low coupling cases, although not higher precision in the singular-value decomposition cutoff or time-step. The relative discrepancy in the asymptotic values between analytical and numerical solutions for $\langle\langle Q^2\rangle\rangle$ in the case of $T=1$ are found to be $\delta Q^2/Q^2 = 0.12\%$ for $\alpha=0.1$, and  $\delta Q^2/Q^2 = 0.06\%$ for $\alpha=1.5$. In the case of $T=0.1$, $\alpha=0.1$, the relative discrepancy is $\delta Q^2/Q^2 = 0.13\%$.

\subsection{Unbiased spin boson model}

We now turn to the spin-boson model with $\Omega\neq 0$, focussing on the unbiased case where $\omega_0=0$. In this context, TEMPO has previously been used to pinpoint the localisation phase transition~\cite{strathearnNature}, which occurs when $T=0$ and at a critical value of the coupling $\alpha$~\cite{Bulla2003,Bulla2005}, and to study non-Markovian dynamics induced by a spatially correlated environments~\cite{Gribben2020}. Here we use it to investigate the non-equilibrium thermodynamics of relaxation over a range of temperatures and coupling strengths. In the following, we take $\Omega=1$, which defines our unit of energy.

\subsubsection{High temperature and weak coupling}

\begin{figure}
\begin{centering}
\includegraphics[scale=0.5]{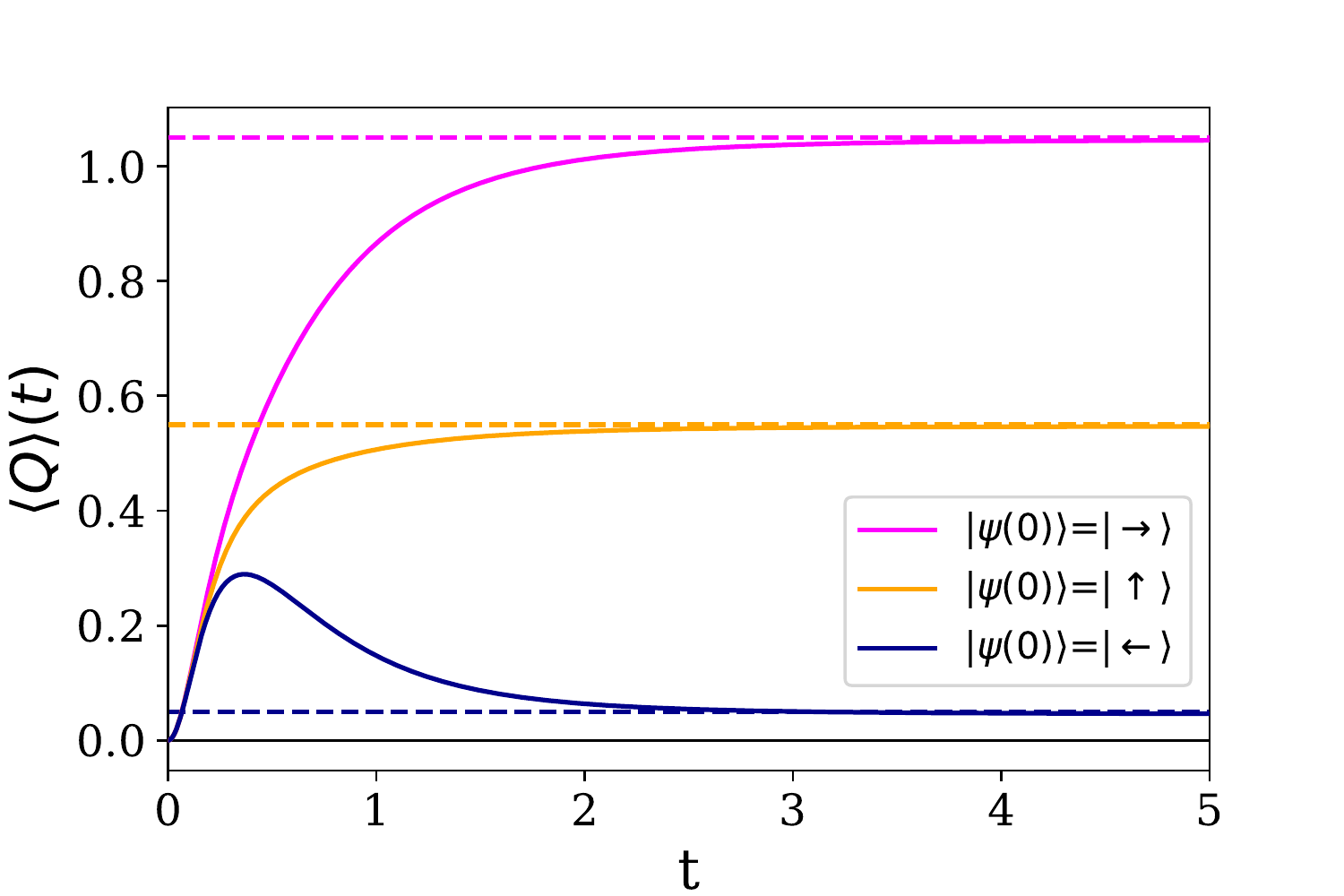}
\includegraphics[scale=0.5]{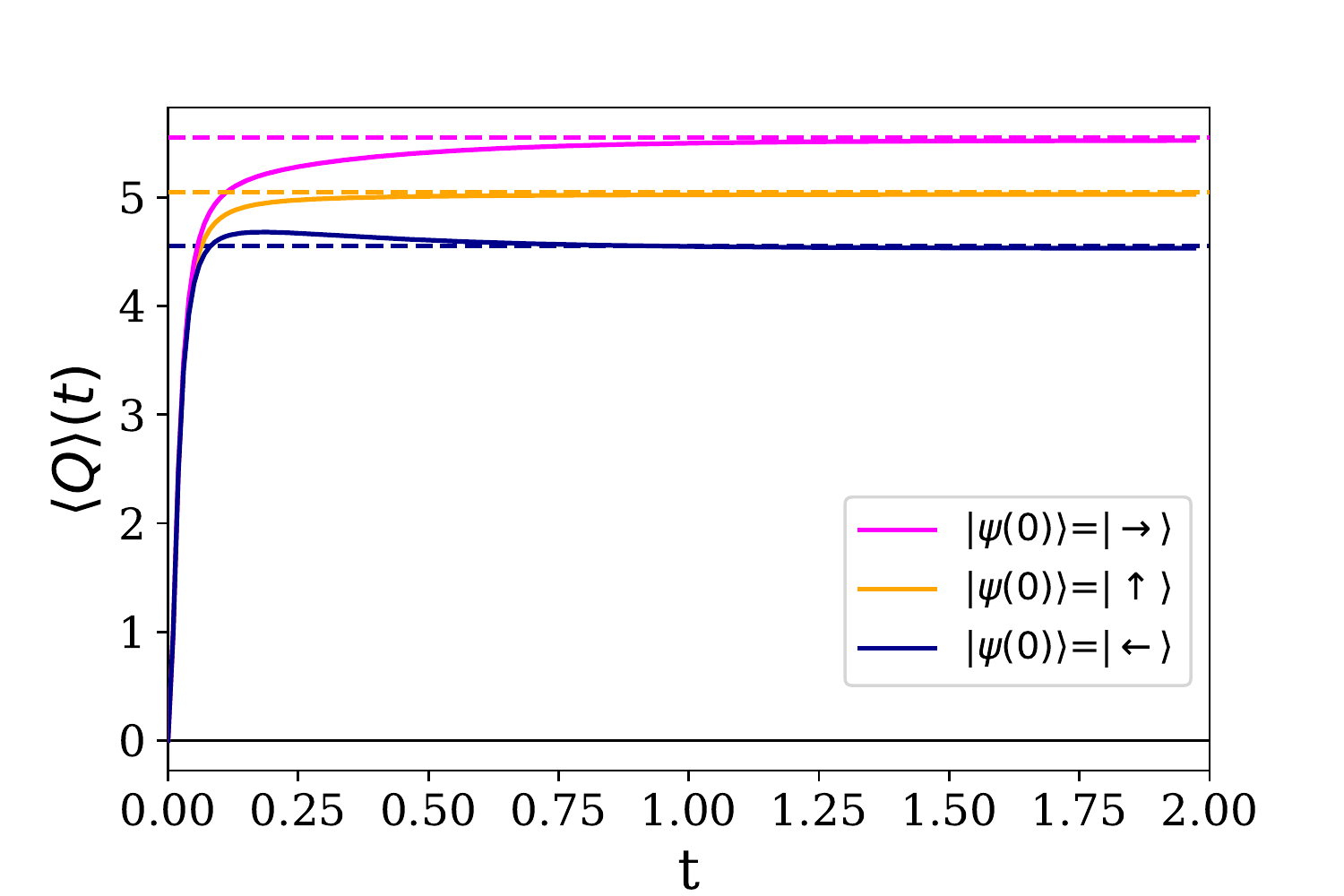}
\par\end{centering}
\caption{Heat transfer for the spin boson model in the high-temperature weak-coupling regime, with bath cut-off $\omega_C=5$ (upper panel) and $\omega_C=50$ (lower panel).  Solid lines: numerical results for the mean heat $\left\langle Q\right\rangle\left(t\right)$ transferred to the bath as a function of time, for three different initial states of the system. Dashed lines: asymptotic approximation for $\langle Q\rangle_\infty$ given by Eq.~\eqref{Q_weak_coupling}. The environment parameters are set to $T=5$ and $\alpha=0.1$. The parameters controlling the numerical accuracy are $K\Delta=5$, $\Delta=0.01$, $p=100$ and the derivative is taken at $u=0.01$ for $\omega_C=5$ and $u=0.001$ for $\omega_C=50$. \label{fig:mean_heat_high_temp}}
\end{figure}

\begin{figure}
\begin{centering}
\includegraphics[scale=0.5]{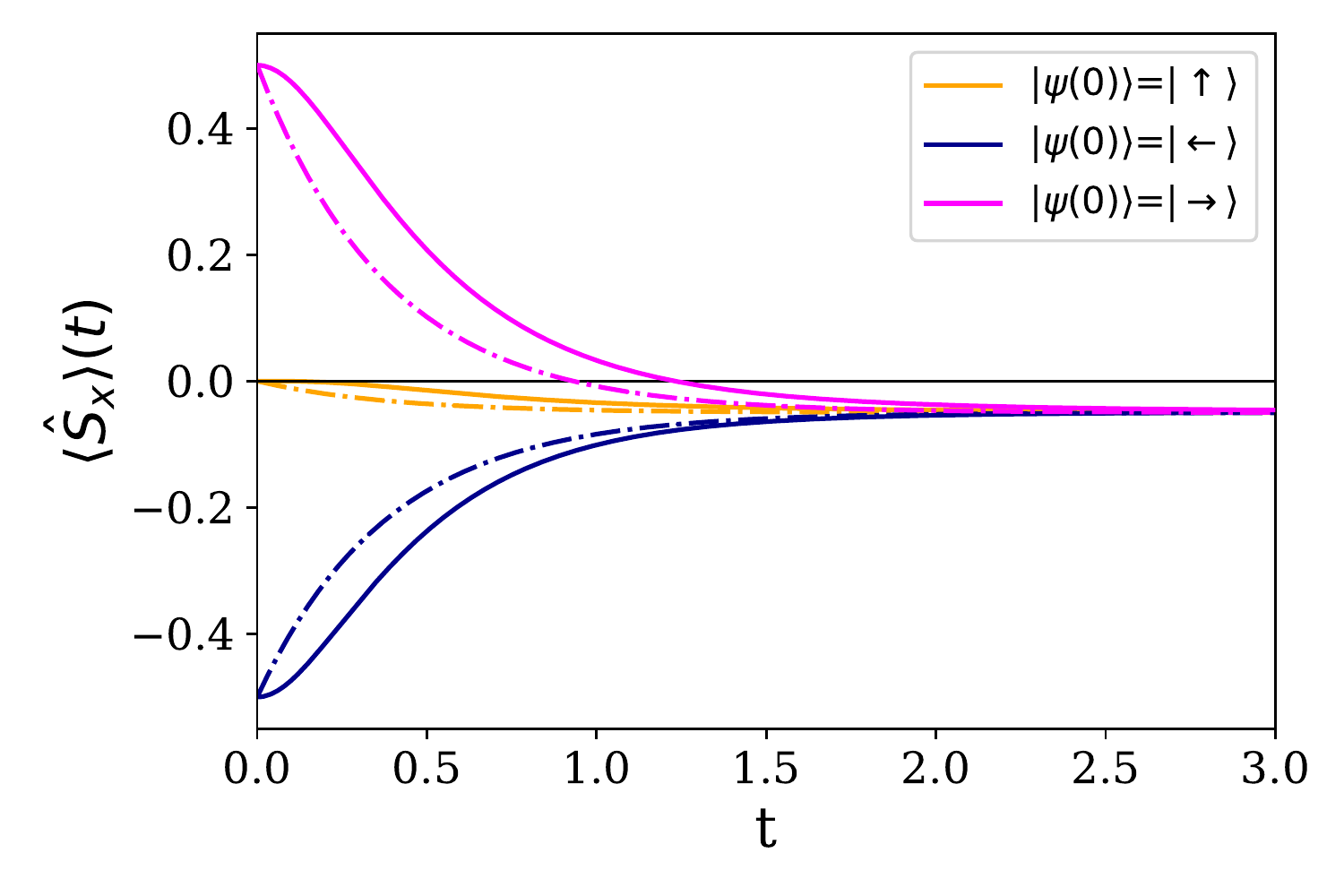}
\par\end{centering}
\caption{Expectation value $\langle \hat{S}_x\rangle(t)$ for the spin boson model at weak coupling and high temperature, for three different initial states of the system. The figure shows a comparison between the numerical results (solid lines) and the results obtained in the Born-Markov and weak-coupling approximation with the same parameters (dash-dotted lines of the same color as the corresponding initial states). The environment parameters are set to $T=5$, $\alpha=0.1$ and $\omega_{C}=5$. The parameters controlling the numerical accuracy are $K\Delta=5$, $\Delta=0.01$, and $p=100$.\label{fig:Sx}}
\end{figure}

We begin by studying the regime of weak coupling and relatively high temperature, with $\alpha=0.1$ and $T=5$. The mean heat transfer is plotted in Fig.~\ref{fig:mean_heat_high_temp} as a function of time, starting from a pure initial state, $\hat{\rho}_S(0) = \ket{\Psi_0}\bra{\Psi_0}$. Specifically, we consider three different initial conditions: $\ket{\Psi_0}\in\{\ket{\leftarrow},\ket{\rightarrow},\ket{\uparrow}\}$, where $\hat{S}_x\ket{\rightarrow}=\tfrac{1}{2}\ket{\rightarrow}$, $\hat{S}_x\ket{\leftarrow}=-\tfrac{1}{2}\ket{\leftarrow}$ and $\hat{S}_z\ket{\uparrow}=\tfrac{1}{2}\ket{\uparrow}$. We also consider two values of the cutoff, $\omega_C=5$ and $\omega_C=50$. Inspection of these results suggests that the heat transfer, $\langle Q\rangle$ is a sum of two contributions. The first contribution is the heat transferred directly from the system as it relaxes to a thermal state $\hat{\rho}^{\rm eq}_S\propto e^{-\beta \hat{H}_S}$. The corresponding change in internal energy will be 
\begin{equation}
    \label{deltaU_weak_coupling}
  \Delta U_\infty = -\frac{\Omega}{2}\tanh\left(\frac{\beta \Omega}{2}\right) - \langle \hat{H}_S\rangle_0.
\end{equation}
The second contribution to the mean heat transfer is associated with switching on the system-bath interaction, and is equivalent to the work done in a cyclic process as discussed in Sec.~\ref{sec:thermodynamics}. If we assume that this contribution is the reorganisation energy, as in the independent boson model, we expect 
\begin{equation}
      \label{Q_weak_coupling}
\langle Q\rangle_\infty = E_r + \frac{\Omega}{2}\tanh\left(\frac{\beta \Omega}{2}\right) + \langle \hat{H}_S\rangle_0.
\end{equation}
This approximation shows near-perfect agreement with the long-time limit of the numerical results, as demonstrated by the dashed lines in Fig.~\ref{fig:mean_heat_high_temp}. Notice that Eq.~\eqref{deltaU_weak_coupling} is independent of the details of the bath spectral density (i.e.~$\alpha$ and $\omega_c$), while $E_r$ does not depend in any way on the spin degrees of freedom. This indicates that, at high temperature and weak coupling, the displacement of the bath modes is not affected by the thermalisation of the spin. Instead, these two processes give rise to independent and additive contributions to the mean heat transfer. 

These distinct modes of heat transfer take place on different time scales. This is  illustrated by the blue lines in both the $\omega_C=5$ and $\omega_C=50$ case of Fig.~\ref{fig:mean_heat_high_temp}, corresponding to the low-energy initial state $\ket{\Psi_0} = \ket{\leftarrow}$. First, heat is transferred to the environment as the system-bath interaction forces the bath modes to rapidly adjust to their new equilibrium. This takes place over a time set by the inverse cutoff, $\omega_C^{-1} \approx 0.2$ for $\omega_C=5$ and $\omega_C^{-1} \approx 0.02$ for $\omega_C=50$. Then, the direction of heat flow reverses as the bath gives up energy in order to bring the spin to thermal equilibrium, which occurs on a slower timescale fixed by the inverse of the thermalisation rate, which can be estimated as $\gamma\approx (\pi/4) J(\Omega)\coth(\beta\Omega/2)$ from standard weak-coupling theories, e.g.~the secular Born-Markov master equation~\cite{breuer}, giving $\gamma^{-1} \approx 0.8$ for $\omega_C=5$ and $\gamma^{-1} \approx 0.65$ for $\omega_C=50$. A comparison between the two different values of $\omega_C$ in Fig.~\ref{fig:mean_heat_high_temp} shows how a larger frequency cut-off determines a shorter timescale for the heat transfer process, for fixed $T$ and $\alpha$. ($E_r$ is ten times larger in the $\omega_C=50$ case, so that the energy due to the displacement of the bath modes dominantes over that due to the spin thermalisation.)

It is worth emphasising that the system-bath interaction energy gives a significant contribution to the heat transfer, even though the system dynamics is very well captured by a Markovian, weak-coupling description. Indeed, for the parameters considered in Fig.~\ref{fig:mean_heat_high_temp} and $\omega_C=5$, the reorganisation energy is comparable to the natural energy scale of the spin, since $E_r = \Omega/2$. Nevertheless, Fig.~\ref{fig:Sx} shows that in this regime the calculated spin dynamics (solid curves) matches the corresponding Born-Markov and weak-coupling approximated problem (dash-dotted curves), within the limits of such an approximation, the coupling strength being set to $\alpha=0.1$. The discrepancy shown in Fig.~\ref{fig:Sx} is $\lesssim 10 \% $.

\subsubsection{Lower temperature and stronger coupling}

\begin{figure}
\begin{centering}
\includegraphics[scale=0.5]{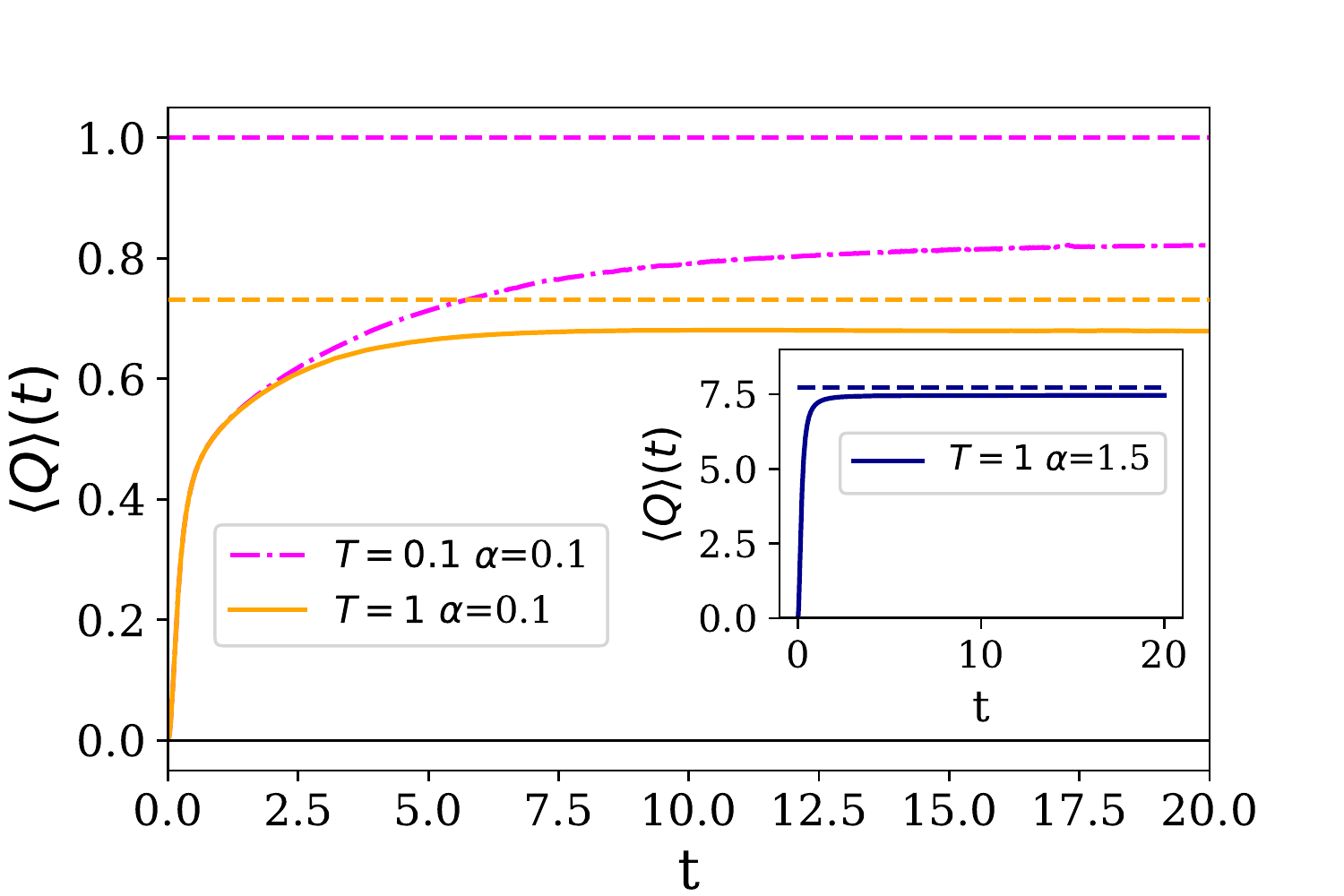}
\par\end{centering}
\caption{Mean heat $\left\langle Q\right\rangle\left(t\right)$ exchanged by the bath for the spin boson model in weak coupling at temperature $T=1$ (solid line) and $T=0.1$ (dash-dotted line), as a function of time, for an initial state of the system set to $\left|\uparrow\right\rangle $. Dashed lines: sum of the energy change in the system and the reorganisation energy of the bath for the corresponding temperatures and coupling strengths. Inset: same plot for temperature $T=1$ and strong coupling. The parameters controlling the numerical accuracy are $K\Delta=5$, $\Delta=0.01$, $p=100$ and the derivative is taken at $u=0.01$. $\omega_{C}=5$ for all the plots.\label{fig:mean_heat_low_temp}}
\end{figure}

We now consider the heat transfer at intermediate and low temperatures. In Fig.~\ref{fig:mean_heat_low_temp} we show the mean heat transfer for temperatures $T=1$ and $T=0.1$, starting from the state $\ket{\Psi_0} = \ket{\uparrow}$. We see the same monotonic relaxation behaviour as was observed at high temperature (cf.~the orange curve in Fig.~\ref{fig:mean_heat_high_temp}), albeit proceeding on a slower timescale as the temperature is reduced. 
\begin{figure}
\begin{centering}
\includegraphics[scale=0.5]{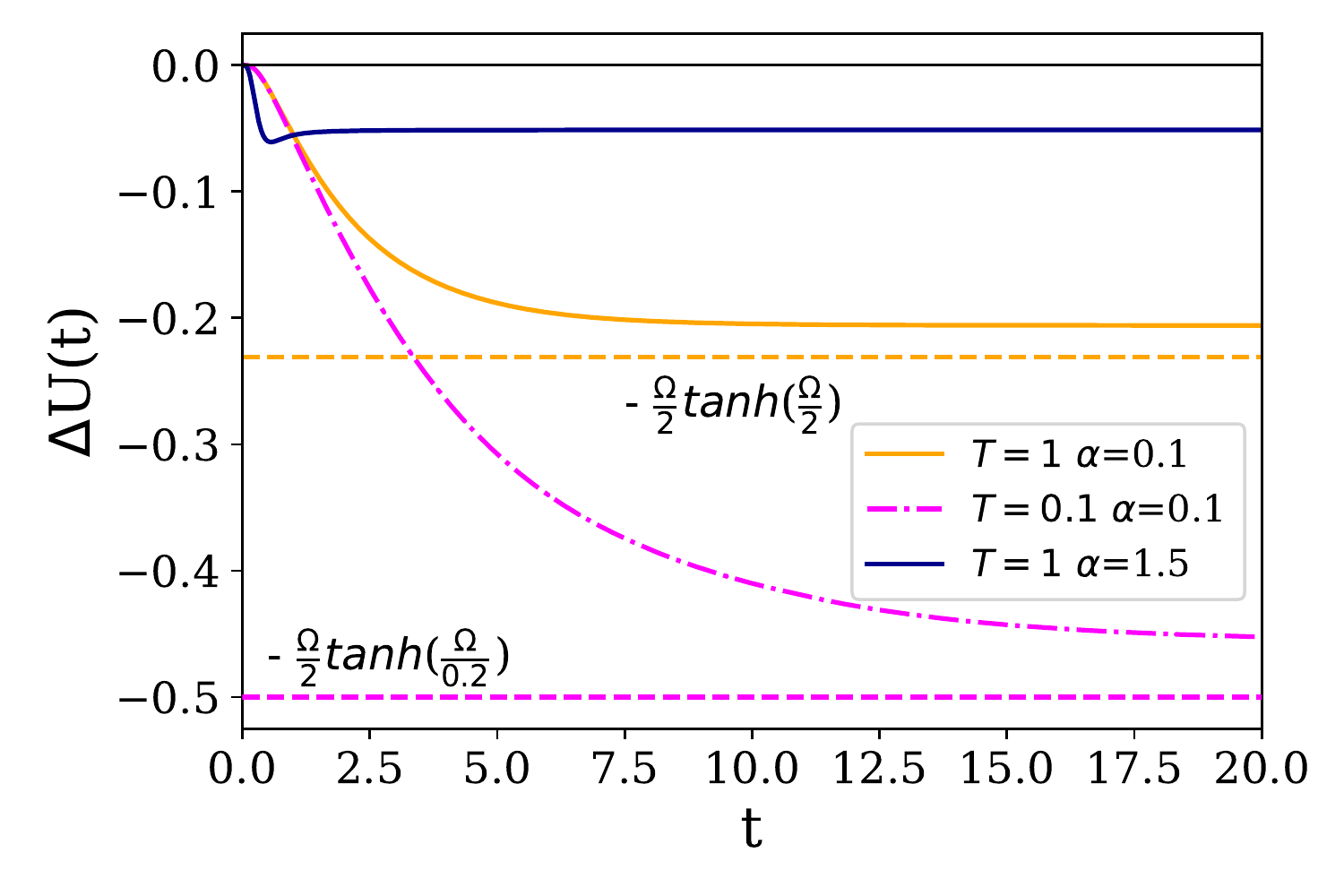}
\par\end{centering}
\caption{\label{fig:system_internal_energy_change}Variation of internal energy
of the system $\Delta U\left(t\right)$ as a function of time for
temperature $T=1$, where the solid blue line is for $\alpha=1.5$ and the solid orange line for $\alpha=0.1$, and temperature $T=0.1$, where the dash-dotted line is for $\alpha=0.1$. Dashed lines: 
total internal energy change of system in the Markovian regime, $-\frac{\Omega}{2}\tanh\left(\frac{\Omega}{2T}\right)$,
for $T=1$ (dashed orange line) and $T=0.1$ (dashed magenta line). The parameters controlling the numerical accuracy are $K\Delta=5$, $\Delta=0.01$ and $p=100$. $\omega_{C}=5$ for all the plots.}
\end{figure}
Outside of the high-temperature limit, the asymptotic value of $\langle Q\rangle$ can no longer be well approximated by Eq.~\eqref{Q_weak_coupling}, shown by the dashed lines in Fig.~\ref{fig:mean_heat_low_temp}. We find that the the spin's internal energy change and the total heat transfer are smaller in magnitude than Eqs.~\eqref{deltaU_weak_coupling} and~\eqref{Q_weak_coupling} predict, as Fig.~\ref{fig:mean_heat_low_temp} and Fig.~\ref{fig:system_internal_energy_change} both show. This demonstrates that the tendency of the spin to minimise its local free energy defined by $\hat{H}_S$ competes with the displacing effect of $\hat{H}_I$ on the bath modes. As a consequence of this interplay, both $\Delta U$ and $\langle W\rangle$ depend non-trivially on system-bath correlations generated during the relaxation process.\\
The effect of the correlations with the bath is indeed to decrease the magnitude of $\Delta U$ with respect to the value $-\frac{\Omega}{2}\tanh\left(\frac{\Omega}{2T}\right)$ predicted by Eq.~\eqref{deltaU_weak_coupling}, and represented in Fig.~\ref{fig:system_internal_energy_change} by the dashed lines. Such discrepancy is starkly greater for stronger coupling.

In order to understand this, we note that at strong system-bath coupling the equilibrium state must be generalised to~\cite{Talkner2020}
\begin{equation}
\label{eq:generalised_equilibrium}
    \hat{\rho}_S^{\rm eq} = \frac{\Tr_B\left[e^{-\beta \hat{H}}\right]}{\Tr\left[e^{-\beta \hat{H}}\right]},
\end{equation}
i.e.~the reduction of a global thermal state. This takes into account correlations with the bath and reduces to the standard form $\hat{\rho}^{\rm eq}_S \propto e^{-\beta  \hat{H}_S }$ in the weak-coupling limit. Assuming that the open quantum system couples to the bath locally in space, the interaction Hamiltonian is a local degree of freedom that is also expected to thermalise, in the sense that
\begin{equation}
\label{eq:H_SB_equilibrium}
    \langle   \hat{H}_{I}  \rangle_\infty = \frac{\Tr\left[  \hat{H}_{I}  e^{-\beta \hat{H}}\right]}{\Tr\left[e^{-\beta \hat{H}}\right]}.
\end{equation}
We emphasise that these thermalisation conditions hold only for local subsystems: they do not imply that the system as a whole attains thermal equilibrium in the long-time limit.

We estimate the effect of system-bath correlations on heat transfer using the variational approach pioneered by Silbey and Harris~\cite{Silbey1984}, which has been successfully applied to understand various static and dynamic properties of the spin-boson model~\cite{McCutcheon2011,Chin2011,Nazir2012}. The method is briefly summarised here with further details given in Appendix~\ref{app:Silbey_Harris}. The basic idea is to express the Hamiltonian in a different basis by applying a unitary transformation that mixes the system and bath degrees of freedom. A judicious choice of transformation --- determined in this case by a variational principle --- leads to a weak effective interaction term $\hat{H}_I'$ in the new basis, even though the bare interaction $\hat{H}_I$ may be strong.

Specifically, the Hamiltonian is diagonalised approximately using the polaron transformation in Eq.~\eqref{polaron_transformation}, with the displacements $\{f_j\}$ interpreted as variational parameters. After the transformation, the Hamiltonian is written as $\hat{P}^\dagger\hat{H} \hat{P} = \hat{H}'_0 + \hat{H}'_I\approx \hat{H}_0'$. Here, $\hat{H}'_0$ is the free Hamiltonian in the variational frame, which is given up to a constant by
\begin{align}
    \label{variational_H0}
    \hat{H}'_0 & = \Omega'\hat{S}_x + \sum_j\omega_j \hat{a}_j^\dagger \hat{a}_j,
\end{align}
where $\Omega'$ is a renormalised tunnelling matrix element to be defined below. The neglected interaction term, $\hat{H}'_I$, describes residual transitions between dressed states of the system and environment and is proportional to the spin tunnelling amplitude $\Omega$. The effect of $\hat{H}'_I$ is made as small as possible by choosing the variational parameters to minimise the Feynman-Bogoliubov upper bound on the free energy; see Appendix~\ref{app:Silbey_Harris} for details. This is achieved by taking $f_j = g_j\phi(\omega_j)$ with
\begin{align}
    \label{phi_def}
    &\phi(\omega) = \left[1 + \frac{\Omega'}{\omega}\tanh \left(\frac{\beta \Omega'}{2}\right)\coth\left(\frac{\beta \omega}{2}\right) \right]^{-1},\\
        \label{B_solution}
    &\Omega'  = \Omega \exp\left[-\frac{1}{2}\int_0^\infty d\omega \frac{J(\omega)}{\omega^2}\phi^2(\omega) \coth\left(\frac{\beta \omega}{2}\right) \right],
    \end{align}
which must be solved self-consistently for $\Omega'$. The heat transfer is then found by approximating $e^{-\beta \hat{H}} \approx \hat{P}e^{-\beta\hat{H}_0'}\hat{P}^\dagger$ in Eqs.~\eqref{eq:generalised_equilibrium} and \eqref{eq:H_SB_equilibrium}, yielding
\begin{equation}
\label{variational_heat}
\langle Q\rangle_\infty = E_r' + \frac{\Omega'}{2} \tanh\left (\frac{\beta \Omega'}{2}\right) + \langle \hat{H}_S\rangle_0.
\end{equation}
This has the same form as Eq.~\eqref{Q_weak_coupling} but with both the tunnelling matrix element $\Omega'$ and reorganisation energy $E_r' = \tfrac{1}{2}\int d\omega J(\omega)\phi(\omega)/\omega$ renormalised.

The variational theory predicts that both the spin tunnelling matrix element and the reorganisation energy are reduced relative to their bare values, since  $\Omega'/\Omega \leq 1$ and $\phi(\omega_j) \leq 1$. Physically, this occurs because the tunnelling between spin states $\ket{\uparrow}\leftrightarrow \ket{\downarrow}$ induced by $\hat{H}_S$ is suppressed by the spin-dependent mode displacements generated by $\hat{H}_I$, which reduce the effective overlap between the two spin states. The equilibrium state emerges from a balance of these two competing effects, which explains why both $\Delta U$ and $\langle Q\rangle$ are reduced at low temperature relative to Eqs.~\eqref{deltaU_weak_coupling} and~\eqref{Q_weak_coupling}.

\begin{figure}
    \centering
    \includegraphics[scale=0.5]{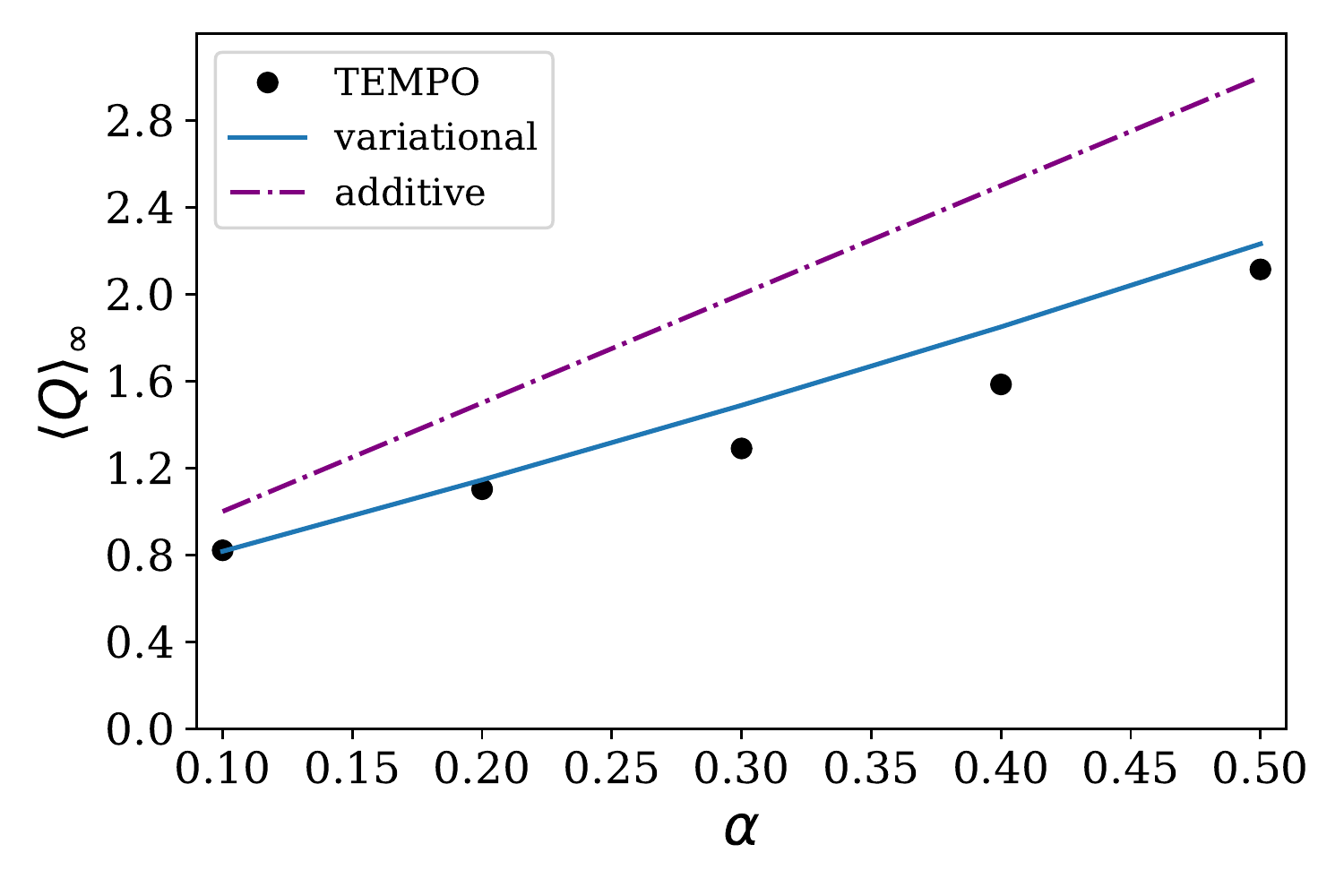}
    \caption{Long-time limit of the heat transfer for the spin boson model as a function of coupling strength, calculated using the path integral (circles), the additive theory (dash-dotted) and the variational method (line), for $T=0.1$.\label{fig:variational_comparison}}
    \label{fig:my_label}
\end{figure}

We show in Fig.~\ref{fig:variational_comparison} that the variational theory gives a good quantitative approximation to the mean heat transfer at low temperature, $T=0.1$, with the best agreement at weak coupling. At higher temperatures on the order of $T=1$ and above, we find that the approximation breaks down completely because the renormalisation of the tunnelling amplitude is overestimated, leading to values $\Omega'\ll \Omega$. This failure is presumably due to the neglect of thermally activated transitions generated by $\hat{H}'_I$, which become relevant at temperatures $\beta \Omega'\lesssim 1$. On the other hand Fig.~\ref{fig:variational_comparison} shows that the additive ansatz given by Eq.~\eqref{Q_weak_coupling} performs worse than the variational theory across all the coupling range.

At very strong coupling, the variational theory performs well at all temperatures. In this regime, strong correlations with the bath lead to an almost maximally mixed equilibrium state of the spin, corresponding to a vanishing tunnelling rate in the variational frame, $\Omega'\to 0$. As a result, the heat transfer for an initial state $\ket{\Psi_0} = \ket{\uparrow}$ reduces to the bare reorganisation energy, $E_r$. This behaviour is shown in the inset of Fig.~\ref{fig:mean_heat_low_temp}, where the solid curve converges to $\langle Q\rangle  \approx E_r$ to a good approximation. The dynamics of the heat transfer is correspondingly fast in this regime since it depends only on the bath cutoff scale, $\omega_C$.

\subsection{Heat fluctuation-dissipation relation in the spin boson model}

\begin{figure}
    \centering
    \includegraphics[scale=0.5]{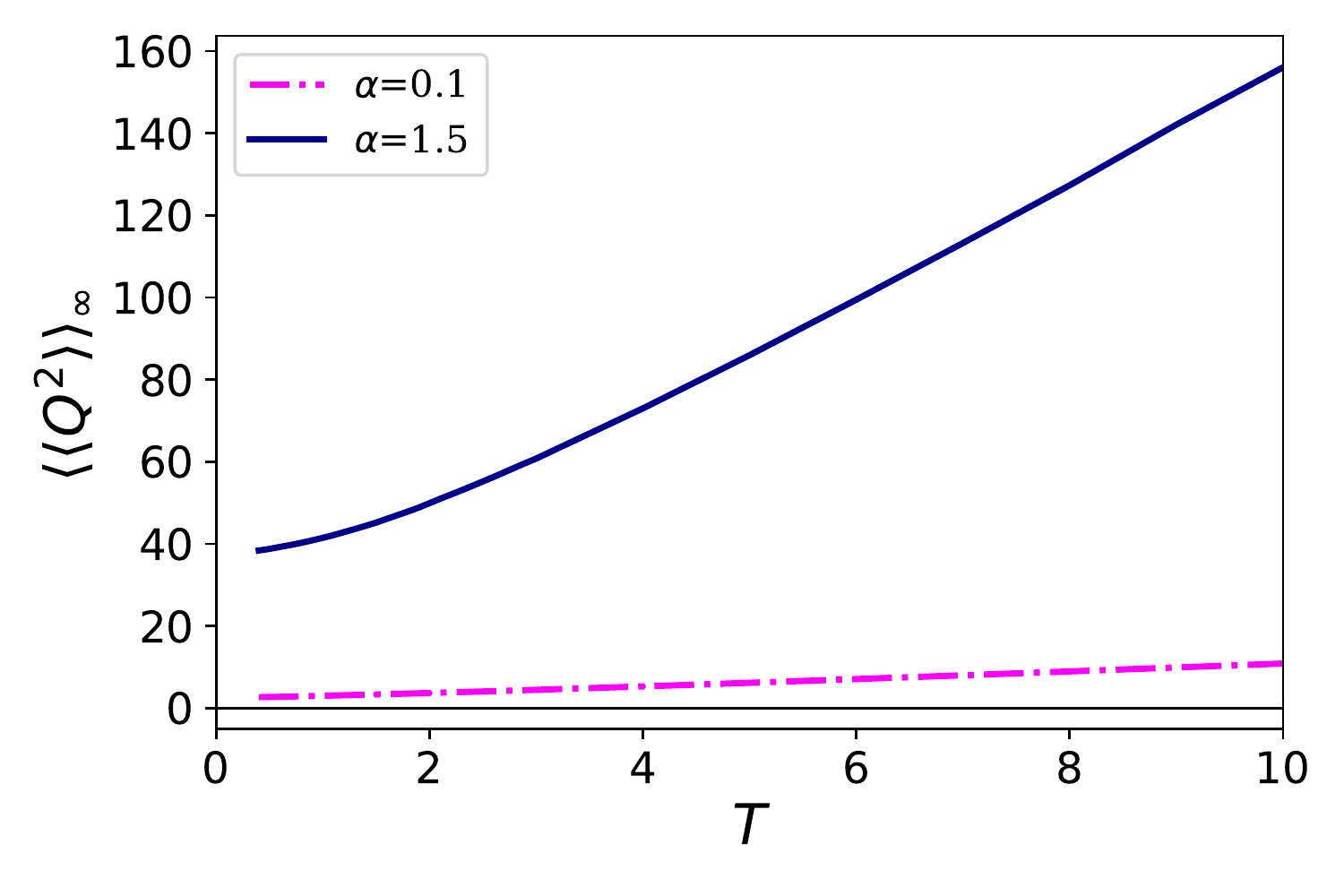}
    \caption{Variance of the heat distribution as a function of temperature in the spin-boson model, for both weak ($\alpha = 0.1$) and strong ($\alpha = 1.5$) coupling starting from the initial state $\ket{\psi(0)} = \ket{\uparrow}$. The parameters are $\omega_{C}=5$, $K\Delta=5$ and $\Delta=0.01$, with $p=100$ and $u=0.005$ for $\alpha=1.5$, and $p=120$ and $u=0.01$ for $\alpha=0.1$.}
    \label{fig:varianceT}
\end{figure}

As a final demonstration of our method, we study the temperature dependence of the heat fluctuations in the spin-boson model. Fig.~\ref{fig:varianceT} shows the asymptotic variance of the heat distribution at long times, starting from the initial state $\ket{\psi(0)} = \ket{\uparrow}$. We see that the fluctuations increase with temperature, and grow approximately linearly with $T$ at high temperature. 

This linear behaviour of $\langle\!\langle Q^2\rangle \!\rangle_\infty$ can be understood as a manifestation of the fluctuation-dissipation relation (FDR) that is well known in the context of non-equilibrium work distributions. If the distribution of work $W$ is Gaussian, the Jarzynski equality directly implies the FDR~\cite{Jarzynski1997} $\langle W\rangle-\Delta F= \beta \langle\!\langle W^2\rangle\!\rangle/2$, where $\Delta F$ is the equilibrium free energy change. In the case of the independent-boson model, heat is identical to work since $\hat{H}_S$ is a conserved quantity, while $\Delta F=0$ because the process is cyclic. It follows that we can write an equivalent FDR for the heat distribution:
\begin{equation}
    \label{fluctuation_dissipation}
    \langle\!\langle Q^2\rangle\!\rangle = 2T \langle Q\rangle.
\end{equation}
At high temperature, this relation holds at all times in the independent-boson model, as can be seen by comparing Eqs.~\eqref{mean_heat_IBM} and~\eqref{variance_IBM} in the limit $\beta\omega_C\ll 1$.

\begin{figure}
    \centering
    \includegraphics[scale=0.5]{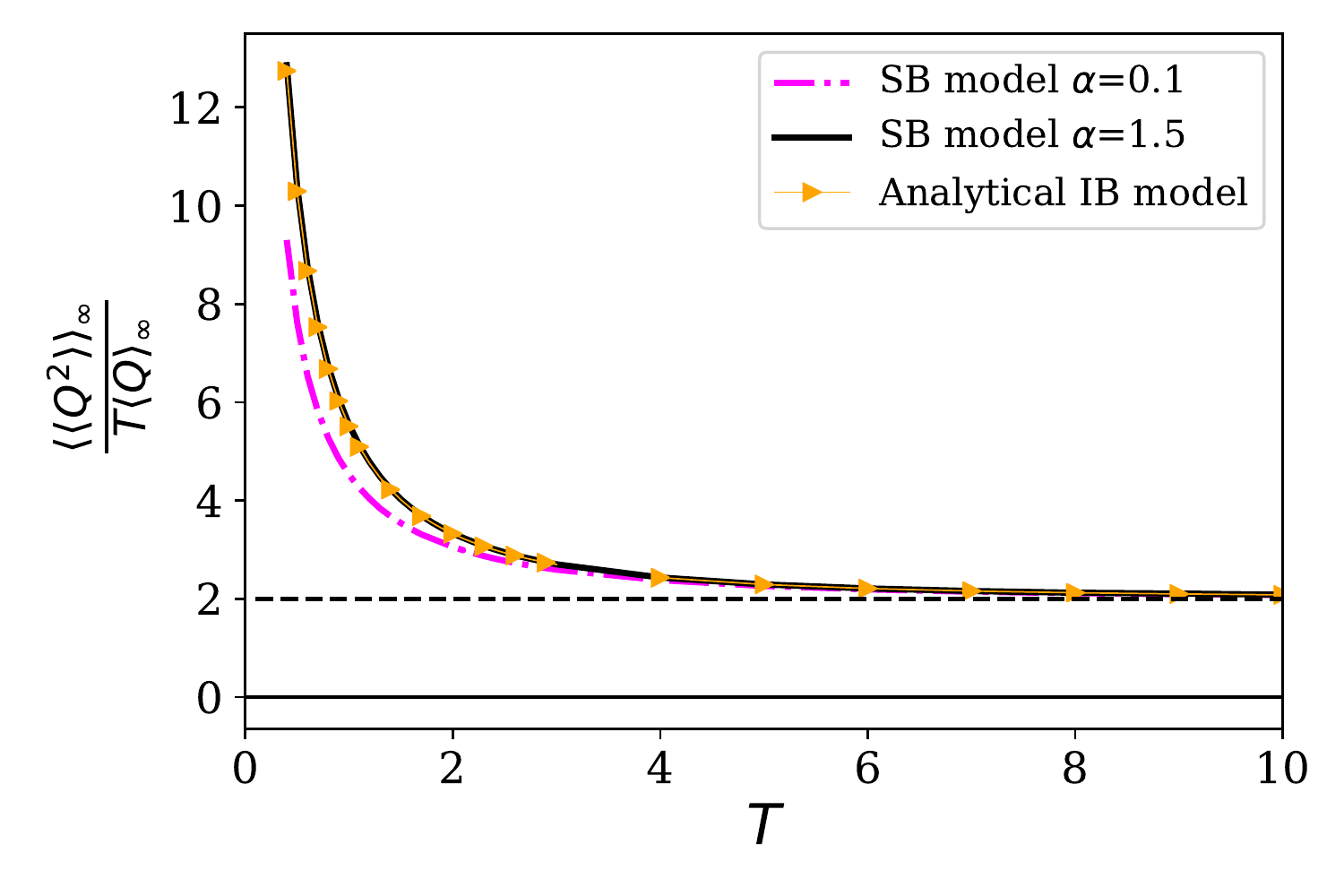}
    \caption{Asymptotic ratio of the variance to the mean heat as a function of temperature, showing the validity of the FDR for $T\gg \omega_C$. Dash-dotted and solid lines are the numerical results for the spin boson model, for the values of $\alpha$ indicated. The figure shows a comparison with the analytical solution for the independent boson model, which is independent of $\alpha$ (triangles). The FDR value of $\langle\!\langle Q^2\rangle\!\rangle_\infty /T\langle Q\rangle_\infty = 2$ is shown by the black dashed line. The parameters are the same as in Fig.~\ref{fig:varianceT}. \label{fig:FDR}}
\end{figure}

In the spin-boson model, we no longer have equality between work and heat since $\Delta U \neq 0$. Nevertheless, we find numerically that the FDR~\eqref{fluctuation_dissipation} approximately holds at high temperatures, $\beta\omega_C\lesssim 1$, as shown in Fig.~\ref{fig:FDR}. This behaviour stems from the fact that the spin's contribution to the heat fluctuations is limited by its finite energy splitting $\Omega$, whereas the contribution of the spin-boson interaction energy can grow arbitrarily large. The heat fluctuations are thus dominated by independent-boson physics at high temperature. For strong coupling, where the spin energy scale $\Omega$ is negligible compared to the reorganisation energy $E_r$, we show in Fig.~\ref{fig:FDR} that the heat fluctuations are essentially identical in the spin-boson and independent-boson models at all temperatures. One limitation encountered in the calculation of the data shown in Fig.~\ref{fig:FDR}, is that the TEMPO algorithm was not able to compute the variance up to equilibrium time for very low temperatures. Indeed, the lowest temperature shown is $T=0.4$. Exploring the validity of the heat FDR in other scenarios, e.g.~multipartite open quantum systems, is an interesting avenue for future work.

\section{Conclusions}
\label{sec:conclusions}

A better understanding of dissipation in open quantum systems is a fundamental goal of quantum thermodynamics as well as being crucial for quantum device engineering. We have shown that this goal can be successfully addressed by an extension of the TEMPO algorithm~\cite{PopovicCode} to evaluate the characteristic function of the heat distribution. We have demonstrated the validity and flexibility of our approach by calculating the mean and variance of the heat transfer in the spin-boson model over a range of temperatures and system-bath coupling strengths. Our results clearly demonstrate the importance of system-environment correlations at low temperatures. Even at high temperature and weak coupling, we find significant contributions to the heat statistics from the system-environment interaction energy that are not captured by the standard weak-coupling master equation. This indicates that system-reservoir interactions are an important source of dissipation that must be accounted for when designing thermodynamic protocols~\cite{GelbwaserKlimovsky2015, Katz2016,Strasberg2016,Newman2017,Perarnau2018,Newman2020}, even in the weak-coupling regime.

Our approach to calculating heat statistics can be extended in several promising directions. It is straightforward to adapt the method to situations with a time-dependent system Hamiltonian, which would enable the characterisation of heat statistics for driven open systems. This problem, which is theoretically challenging even for Markovian environments outside of the slow-driving regime, has numerous applications in quantum control, such as quantum information processing~\cite{Cimini2020} and erasure~\cite{Miller2020}, enhanced engine cycles through thermodynamic shortcuts~\cite{Dann2020,Pancotti2020}, and tailored quantum light sources~\cite{Murphy2019,IlesSmith2019}. It is also possible to incorporate multiple baths within our framework by combining the corresponding influence functionals together. This would allow the study of the full counting statistics of quantum heat transport in non-equilibrium steady states~\cite{Kilgour2019}, including highly non-Markovian regimes. In general, we expect that the method presented here will facilitate further research into the non-equilibrium quantum thermodynamics of strongly coupled open systems.

\section{Acknowledgements}
\label{sec:acknowledgements}

We are grateful to Dvira Segal for the illuminating discussions on heat statistics and path integral methods. We also thank Peter Kirton, Jonathan Keeling and Dominic Gribben for their input and suggestions. We acknowledge funding from the European Research Council under the European Union's Horizon 2020 research and innovation program (ODYSSEY grant agreement No. 758403). JG is grateful for support from a SFI-Royal Society University Research Fellowship. We also acknowledge support from EPSRC grant EP/T014032/1. We acknowledge the DJEI/DES/SFI/HEA Irish Centre for High-End Computing (ICHEC) for the provision of computational facilities and support, project TCPHY118B. Some calculations were performed on the Lonsdale cluster maintained by the Trinity Centre for High Performance Computing. This cluster was funded through grants from Science Foundation Ireland. AS acknowledges support the Australian Research Council Centres of Excellence for    Engineered Quantum Systems (EQUS, CE170100009).

\bibliographystyle{apsrev4-1}
\bibliography{bibliography}{}

\appendix

\section{Superoperator methods for the calculation of the modified influence functional}
\label{app:SU_methods}
Let $\hat{K}$ be a superoperator acting on the space of bounded operators
$\mathcal{B}\left(\mathcal{H}\right)$ on a Hilbert space $\mathcal{H}$.
Then two superoperators labelled left, $\hat{K}^{L}$, and right,
$\hat{K}^{R}$, can be defined by their actions on a density matrix
operator $\rho$ $\in$ $\mathcal{B}\left(\mathcal{H}\right)$:
\begin{align}
\hat{K}^{L}\rho & =\hat{K}\rho\label{eq:Left},\\
\hat{K}^{R}\rho & =\rho\hat{K}\label{eq:Right}.
\end{align}
Similarly, the superoperators $\hat{K}^{+}$ and $\hat{K}^{-}$ can
be defined as
\begin{align}
\hat{K}^{+}\rho & =\hat{K}^{L}\rho+\hat{K}^{R}\rho=\left\{ \hat{K},\rho\right\}, \label{eq:plus}\\
\hat{K}^{-}\rho & =\hat{K}^{L}\rho-\hat{K}^{R}\rho=\left[\hat{K},\rho\right],\label{eq:Minus}
\end{align}
with inverse transformations
\begin{align}
\hat{K}^{L}\rho & =\frac{1}{2}\left(\hat{K}^{+}+\hat{K}^{-}\right)\rho=\frac{1}{2}\left[\hat{K},\rho\right]+\frac{1}{2}\left\{ \hat{K},\rho\right\} ,\label{eq:invl}\\
\hat{K}^{R}\rho & =\frac{1}{2}\left(\hat{K}^{+}-\hat{K}^{-}\right)\rho=\frac{1}{2}\left\{ \hat{K},\rho\right\} -\frac{1}{2}\left[\hat{K},\rho\right].\label{eq:invr}
\end{align}
In order the evaluate the exponential in Eq.~\eqref{eq:mod_IF},
it is convenient to rewrite the modified interaction Hamiltonian $\tilde{H}_{I}(t,u)=e^{i\hat{H}_{B}u/2}\tilde{H}_{I}(t)e^{-i\hat{H}_{B}u/2}$ as follows:
\begin{align}
\tilde{H_{I}}\left(t,u\right)&=\hat{S}_{z}\sum_{j}g_{j}\cos\left(\frac{u}{2}\omega_{j}\right)\left(a_{j}e^{-i\omega_{j}t}+a_{j}^{\dagger}e^{i\omega_{j}t}\right)+\nonumber \\
&-i\hat{S}_{z}\sum_{j}g_{j}\sin\left(\frac{u}{2}\omega_{j}\right)\left(a_{j}e^{-i\omega_{j}t}-a_{j}^{\dagger}e^{i\omega_{j}t}\right)\label{eq:interactionH},
\end{align}
and defining
\begin{align}
B_{1}\left(t,u\right) & =\sum_{j}g_{j}\cos\left(\frac{u}{2}\omega_{j}\right)\left(a_{j}e^{-i\omega_{j}t}+a_{j}^{\dagger}e^{i\omega_{j}t}\right),\label{eq:b1}\\
B_{2}\left(t,u\right) & =-i\sum_{j}g_{j}\sin\left(\frac{u}{2}\omega_{j}\right)\left(a_{j}e^{-i\omega_{j}t}-a_{j}^{\dagger}e^{i\omega_{j}t}\right),\label{eq:b2}
\end{align}
we have
\begin{equation}
\tilde{H_{I}}\left(t,u\right)=\hat{S}_{z}B_{1}\left(t,u\right)+\hat{S}_{z}B_{2}\left(t,u\right).\label{eq:sum}
\end{equation}
The interaction Hamiltonian has thus been divided into the sum of the two Hamiltonians
\begin{align}
H_{I,1}\left(t,u\right) & =\hat{S}_{z}B_{1}\left(t,u\right),\label{eq:H1}\\
H_{I,2}\left(t,u\right) & =\hat{S}_{z}B_{2}\left(t,u\right).\label{eq:H2}
\end{align}
We note that given the cosine and sine functions in the interaction
parts dependent on the counting field, equations (\ref{eq:b1}) and
(\ref{eq:b2}), it holds that $H_{I,1}\left(t,-u\right)=H_{I,1}\left(t,u\right)$
and $H_{I,2}\left(t,-u\right)=-H_{I,2}\left(t,u\right)$. In light
of this new notation, the Liouvillian operator defined in Eq.~\eqref{modified_Liouvillian}
is 
\begin{equation}
\mathcal{L}_{I}\left(t,u\right)=-i\left(H_{I,1}^{-}\left(t,u\right)+H_{I,2}^{+}\left(t,u\right)\right)\label{eq:liouvillian},
\end{equation}
where we have used Eq.~\eqref{eq:plus} and Eq.~\eqref{eq:Minus}. The exponent of the modified influence functional in Eq.~\eqref{eq:mod_IF} can then be written as
\begin{align}
&\left\langle \mathcal{L}_{I}\left(t',u\right)\mathcal{L}_{I}\left(t'',u\right)\right\rangle _{B} = \nonumber\\
&-\left\langle H_{I,1}^{-}\left(t',u\right)H_{I,1}^{-}\left(t'',u\right)\right\rangle _{B}-\left\langle H_{I,2}^{+}\left(t',u\right)H_{I,2}^{+}\left(t'',u\right)\right\rangle _{B}+ \nonumber \\
 & -\left\langle H_{I,1}^{-}\left(t',u\right)H_{I,2}^{+}\left(t'',u\right)\right\rangle _{B}-\left\langle H_{I,2}^{+}\left(t',u\right)H_{I,1}^{-}\left(t'',u\right)\right\rangle _{B}.
\label{eq:term1}
\end{align}
Using the decomposition defined in Eq.~\eqref{eq:H1} and Eq.~\eqref{eq:H2},
and rules (\ref{eq:plus} - \ref{eq:invr}), we can write $H_{I,j}^{\pm}\left(t,u\right)=(\hat{S}_{z}B_{j}\left(t,u\right))^{L}\pm (\hat{S}_{z}B_{j}\left(t,u\right))^{R}$, with $j=1,2$. Applying the properties
$\left(AB\right)^{L}=A^{L}B^{L}$ and $\left(AB\right)^{R}=A^{R}B^{R}$, it is possible to separate the superoperator acting on the system operators from those acting on the reservoir operators. Each term in Eq.~\eqref{eq:term1} can then be calculated explicitly:
\begin{widetext}
\begin{align}
\left\langle H_{I,1}^{-}\left(t',u\right)H_{I,2}^{+}\left(t'',u\right)\right\rangle _{B}&=\frac{1}{4}\hat{S}_{z}^{-}\left(\hat{S}_{z}^{+}\left\langle B_{1}^{+}\left(t',u\right)B_{2}^{+}\left(t'',u\right)\right\rangle _{B}+\hat{S}_{z}^{-}\left\langle B_{1}^{+}\left(t',u\right)B_{2}^{-}\left(t'',u\right)\right\rangle _{B}\right)\label{eq:nm},\\
\left\langle H_{I,2}^{+}\left(t',u\right)H_{I,1}^{-}\left(t'',u\right)\right\rangle _{B} & =\frac{1}{4}\hat{S}_{z}^{+}\left(\hat{S}_{z}^{+}\left\langle B_{2}^{+}\left(t',u\right)B_{1}^{-}\left(t'',u\right)\right\rangle _{B}+\hat{S}_{z}^{-}\left\langle B_{2}^{+}\left(t',u\right)B_{1}^{+}\left(t'',u\right)\right\rangle _{B}\right)\label{eq:nm2},\\
\left\langle H_{I,1}^{-}\left(t',u\right)H_{I,1}^{-}\left(t'',u\right)\right\rangle _{B} & =\frac{1}{4}\hat{S}_{z}^{-}\left(\hat{S}_{z}^{+}\left\langle B_{1}^{+}\left(t',u\right)B_{1}^{-}\left(t'',u\right)\right\rangle _{B}+\hat{S}_{z}^{-}\left\langle B_{1}^{+}\left(t',u\right)B_{1}^{+}\left(t'',u\right)\right\rangle _{B}\right)\label{eq:nm3},\\
\left\langle H_{I,2}^{+}\left(t',u\right)H_{I,2}^{+}\left(t'',u\right)\right\rangle _{B} & =\frac{1}{4}\hat{S}_{z}^{+}\left(\hat{S}_{z}^{+}\left\langle B_{2}^{+}\left(t',u\right)B_{2}^{+}\left(t'',u\right)\right\rangle _{B}+\hat{S}_{z}^{-}\left\langle B_{2}^{+}\left(t',u\right)B_{2}^{-}\left(t'',u\right)\right\rangle _{B}\right)\label{eq:nm4}.
\end{align}
\end{widetext}
It can be noted that given the definition in Eq.~\eqref{reservoir_average}, for any two superoperators
$\alpha$ and $\beta$, it holds that $\left\langle \alpha^{-}\beta^{\pm}\right\rangle _{B}=Tr_{B}\left[\left[\alpha,\beta^{\pm}\tilde{\rho}_{B}\left(0\right)\right]\right]=0$.
Therefore the superoperators $B_{1}^{-}\left(t',u\right)$ and $B_{2}^{-}\left(t',u\right)$
in Eqs. ~\eqref{eq:nm} - ~\eqref{eq:nm4} produce null terms, $B\left(t,u\right)$
being the only operator that contains degrees of freedom of the bath $B$. Evaluating the non-null correlations $\left\langle B_{m}^{+}\left(t',u\right)B_{n}^{+}\left(t'',u\right)\right\rangle _{B}$
and $\left\langle B_{m}^{+}\left(t',u\right)B_{n}^{-}\left(t'',u\right)\right\rangle _{B}$, with $m,n=1,2$, is straightforward when using the
definitions in Eq.~\eqref{eq:b1} and Eq.~\eqref{eq:b2} and the properties
of the bosonic operators. The results are
\begin{align}
\left\langle B_{m}^{+}\left(t',u\right)B_{n}^{+}\left(u,t''\right)\right\rangle _{B}&=\left(-1\right)^{m}4Re\left[\mathcal{C}\left(t',t'',u\right)\right]\label{eq:corr1},\\
\left\langle B_{m}^{+}\left(t',u\right)B_{n}^{-}\left(t'',u\right)\right\rangle _{B}&=\left(-1\right)^{m}4iIm\left[\mathcal{C}\left(t',t'',u\right)\right]\label{eq:corr2},
\end{align}
for $m\neq n$, $m,n=1,2$, and
\begin{align}
\left\langle B_{m}^{+}\left(t',u\right)B_{m}^{+}\left(t'',u\right)\right\rangle _{B}&=4Re\left[\mathcal{A}_{m}\left(t',t'',u\right)\right]\label{eq:corr3},\\
\left\langle B_{m}^{+}\left(t',u\right)B_{m}^{-}\left(t'',u\right)\right\rangle _{B}&=4iIm\left[\mathcal{A}_{m}\left(t',t'',u\right)\right]\label{eq:corr4},
\end{align}
where
\begin{align}
\mathcal{C}\left(t',t'',u\right)&=i\int_{0}^{\infty}d\omega J\left(\omega\right)\cos\left(\frac{u}{2}\omega\right)\sin\left(\frac{u}{2}\omega\right) \nonumber\\
& \times \frac{\sinh\left(i\omega\left(t'-t''\right)-\beta_{0}\omega/2\right)}{\sinh\left(\beta_{0}\omega/2\right)},\label{eq:correlation_continuous}
\end{align}
\begin{align}
\mathcal{A}_{1}\left(t',t'',u\right)&=\int_{0}^{\infty}d\omega J\left(\omega\right)\cos^{2}\left(\frac{u}{2}\omega\right)\nonumber \\
& \times \frac{\cosh\left(i\omega\left(t'-t''\right)-\beta_{0}\omega/2\right)}{\sinh\left(\beta_{0}\omega/2\right)}\label{eq:a1},
\end{align}
\begin{align}
\mathcal{A}_{2}\left(t',t'',u\right)&=\int_{0}^{\infty}d\omega J\left(\omega\right)\sin^{2}\left(\frac{u}{2}\omega\right)\nonumber\\
&\times \frac{\cosh\left(i\omega\left(t'-t''\right)-\beta_{0}\omega/2\right)}{\sinh\left(\beta_{0}\omega/2\right)}\label{eq:a2}.
\end{align}
Here $\beta_{0}$ is the inverse temperature of the bath at the initial time, and $J\left(\omega\right)$ its spectral density. The correlation functions introduced in Eqs. (\ref{eq:cont_etaC} - \ref{eq:cont_etaA2}) are calculated as $\eta^{\mathcal{\alpha}}\left(t,u\right)=\int_{0}^{t}dt'\int_{0}^{t'}dt''\alpha\left(t',t'',u\right)$, with $\alpha=\mathcal{C}$,$\mathcal{A}_{1}$, $\mathcal{A}_{2}$. The exponent in the influence functional defined in Eq.~\eqref{eq:mod_IF} is calculated from its form in Eq.~\eqref{eq:term1} by using the analytical results obtained in (\ref{eq:corr1} - \ref{eq:corr4}).

\section{Characteristic function for the independent boson model}
\label{app:chi_IBM}

In this Appendix, we detail the calculation of the characteristic function for the independent boson model, defined by Eqs.~\eqref{eq:systemHam}--\eqref{eq:interHam} with $\Omega=0$. The Hamiltonian is diagonalised by the transformation
\begin{equation}
    \label{polaron_transformation_appendix}
    \hat{P} = \exp \left[\hat{S}_z\sum_j \frac{g_j}{\omega_j} \left(\hat{a}_j - \hat{a}_j^\dagger\right)\right],
\end{equation}
leading to $\hat{P}^\dagger H\hat{P} = \hat{H}_0 - \tfrac{1}{2}E_r$, where $\hat{H}_0 =  \hat{H}_S  +  \hat{H}_B $ is the free Hamiltonian. The reorganisation energy shift proportional to $E_r$ leads to an irrelevant global phase that will be neglected henceforth. Using this transformation, we write the unitary time evolution operator as 
\begin{align}
    \hat{U}(t) & = \hat{P} e^{-i\hat{H}_0 t}\hat{P}^\dagger \\ 
    & = e^{-i \hat{H}_0 t} \left( e^{i \hat{H}_0 t/2}\tilde{P}(t/2) \tilde{P}^\dagger(-t/2) e^{-i \hat{H}_0 t/2}\right),
\end{align}
where the tilde denotes an operator in the interaction picture with respect to $\hat{H}_0$, i.e. 
\begin{align}
 \tilde{P}(t) & = e^{i \hat{H}_0t} \hat{P} e^{-i \hat{H}_0t} \notag \\
 &= \exp \left[\hat{S}_z\sum_j \frac{g_j}{\omega_j}\left( e^{-i \omega_j t}\hat{a}_j - e^{i\omega_jt} \hat{a}_j^\dagger\right)\right].
\end{align}
Using the Baker-Campbell-Hausdorff formula, $e^A e^B = \exp(A+B+\tfrac{1}{2}[A,B]+\ldots)$, and neglecting an irrelevant phase factor, we obtain $ \hat{U}(t)= \hat{U}_0(t) \hat{U}_I(t)$, where $\hat{U}_0(t) = e^{-i \hat{H}_0t}$ is the free propagator and
\begin{equation}
    \label{interaction_propagator_IBM}
    \hat{U}_I(t) = \exp \left[ 2\hat{S}_z \sum_j \left(\alpha_j(t)\hat{a}_j^\dagger - \alpha_j^*(t) \hat{a}_j\right) \right]
\end{equation}
is the interaction-picture propagator, which describes a spin-dependent displacement for each mode of magnitude
\begin{equation}
    \label{displacement}
    \alpha_j(t) = \frac{g_j}{2\omega_j}\left(1-e^{i\omega_j t}\right).
\end{equation}
Note that Eq.~\eqref{interaction_propagator_IBM} can also be derived directly using the Magnus expansion of the time-ordered exponential~\cite{Blanes2009}.

We now plug our expression for $\hat{U}(t)$ into Eq.~\eqref{eq:char_fun_2} to obtain $\chi(u) = \langle \bar{V}_{-u}^\dagger(t) \bar{V}_{u}(t)\rangle_0$,taking the average at time $t=0$, and where $\bar{V}_{u}(t) = e^{i u  \hat{H}_B /2} \hat{U}_I(t)  e^{-i u  \hat{H}_B /2}$ is the modified interaction-picture evolution operator, given explicitly by
\begin{align}
    \bar{V}_{u}(t) & = \ket{\uparrow}\bra{\uparrow}\otimes\prod_j \hat{D}\left(\alpha_je^{i\omega_j u/2}\right) \notag \\ & \quad + \ket{\downarrow}\bra{\downarrow}\otimes\prod_j \hat{D}^\dagger\left(\alpha_je^{i\omega_j u/2}\right),
\end{align}
with $\hat{D}(x) = e^{x \hat{a}^\dagger - x^* \hat{a}}$ the displacement operator for each bosonic mode. We therefore obtain $\chi(u) = p_\uparrow \chi_\uparrow(u) + p_\downarrow \chi_\downarrow(u)$, where $p_\uparrow = \bra{\uparrow}\hat{\rho}_S(0)\ket{\uparrow}$ and $p_\downarrow = \bra{\downarrow}\hat{\rho}_S(0)\ket{\downarrow}$ denote the initial spin occupations and 
\begin{align}
\chi_\uparrow(u) & = \prod_j\left\langle \hat{D}^\dagger\left(\alpha_j e^{-i\omega_j u/2}\right)\hat{D}\left(\alpha_j e^{i\omega_j u/2}\right) \right\rangle_0,\\
\chi_\downarrow(u) & = \prod_j\left\langle \hat{D}\left(\alpha_j e^{-i\omega_j u/2}\right)\hat{D}^\dagger\left(\alpha_j e^{i\omega_j u/2}\right) \right\rangle_0.
\end{align}
These can be evaluated using the property $\hat{D}(x)\hat{D}(y) = e^{i{\rm Im}( xy^*)}\hat{D}(x+y)$ and the thermal average $\langle \hat{D}(x)\rangle = \exp[-\tfrac{1}{2}|x|^2\coth(\beta\omega/2)]$. We find that $\chi_\uparrow(u) = \chi_\downarrow(u)$ and therefore $\chi(u)$ is independent of the spin populations. The final result for $\chi(u)$ is quoted in Eq.~\eqref{chi_IBM}, from which the $n$th cumulant of the heat distribution can be derived via the formula
\begin{equation}
    \label{heat_cumulant_generation}
    \langle\! \langle Q^n\rangle \!\rangle = (-i)^n \left.\frac{d^n}{du^n}\ln \chi(u)\right\rvert_{u=0}.
\end{equation}{}
Explicitly, we obtain 
\begin{align}
\langle\! \langle Q^{2l-1}\rangle \!\rangle & = \frac{1}{2}\!\int_0^\infty \!\!d\omega\, J(\omega) \omega^{2l-3}\left[1-\cos(\omega t)\right] , \\
\langle\! \langle Q^{2l}\rangle \!\rangle & = \frac{1}{2}\!\int_0^\infty \!\!d\omega\, J(\omega) \omega^{2l-2}\left[1-\cos(\omega t)\right] \coth\left(\frac{\beta\omega}{2}\right)\label{second_cumulant_IBM},
\end{align}
for integers $l>0$. We see that all cumulants are positive and only the even cumulants depend on temperature.

\section{Numerical efficiency of the modified TEMPO method}
\label{app:convergence_parameters}

\subsection{Numerical derivative and counting field value}

\begin{figure}
    \centering
    \includegraphics[scale=0.5]{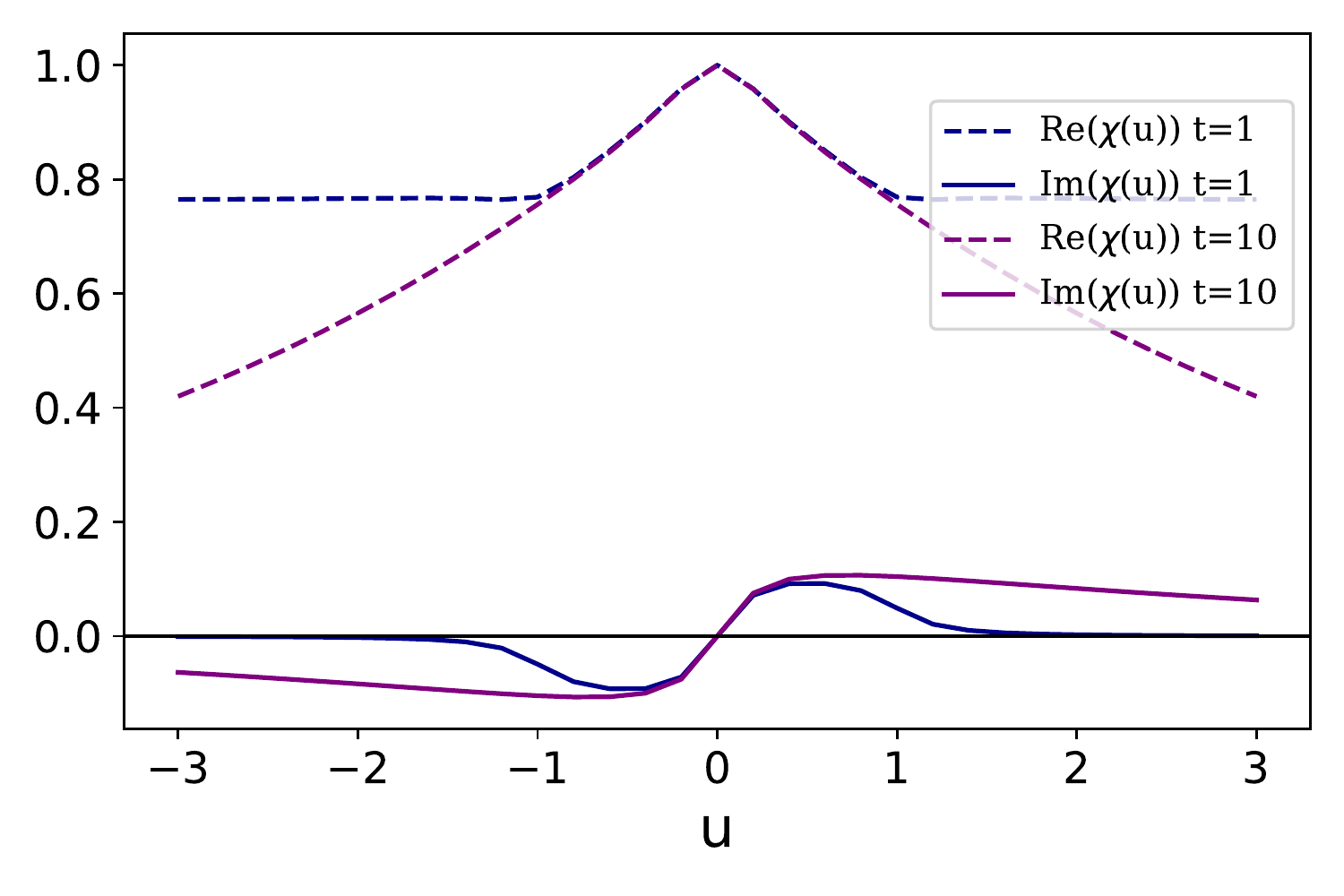}
    \includegraphics[scale=0.5]{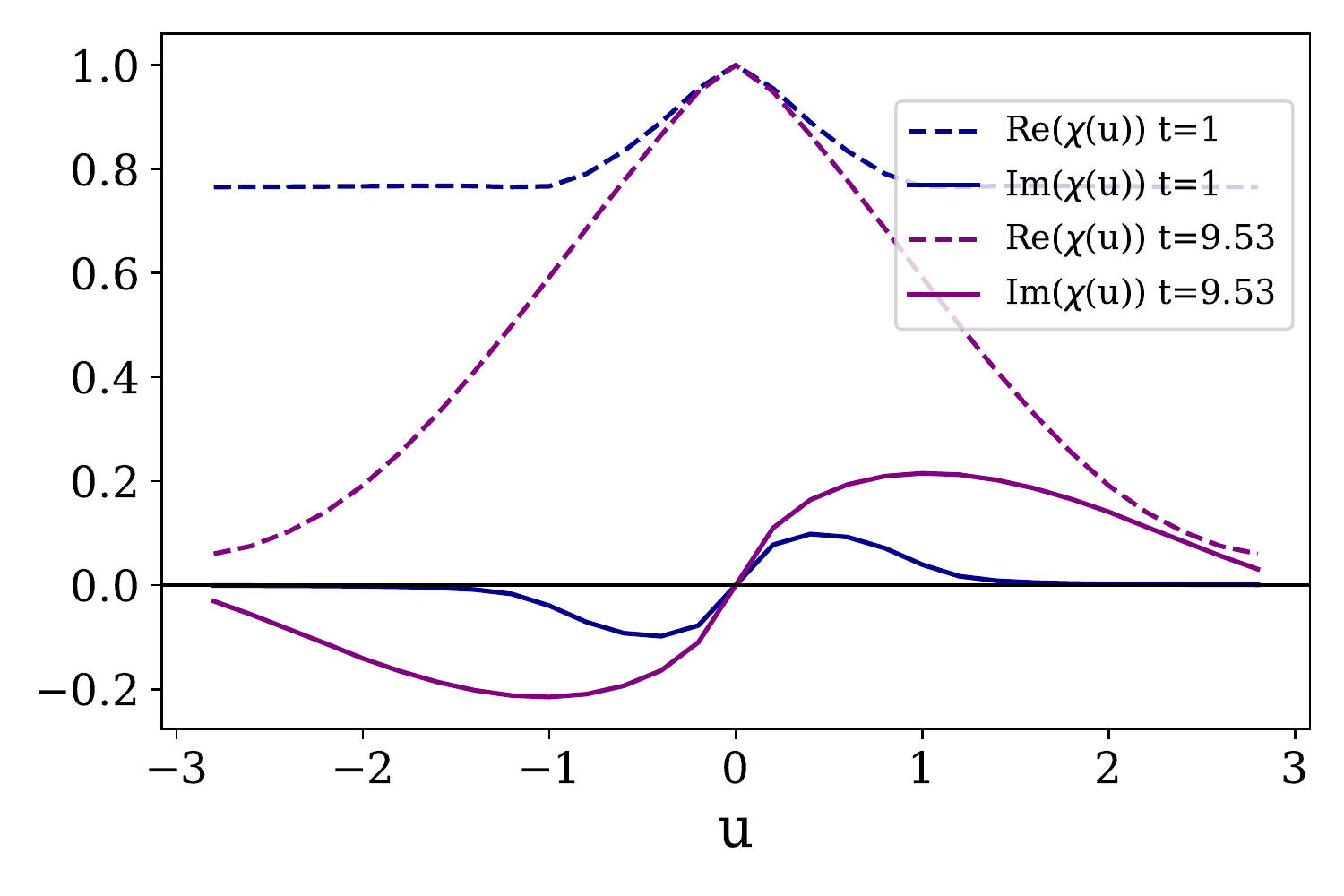}
    \caption{Upper figure: independent-boson model. Lower figure: spin-boson model. Real (dashed line) and imaginary (solid line) parts of the characteristic function, as a function of the counting field parameter $u$. $\chi(u)$ is evaluated for both a small time $t=1$ (blue) and equilibrium times $t=10$ for the IB model and $t=9.53$ for the SB model (purple). The temperature is set to $T=1$ and the coupling strength to $\alpha=0.1$. The parameters controlling the numerical accuracy are $\omega_{C}=5$, $K\Delta=5$, $\Delta=0.01$, and $p=100$. The sampling of the function is taken at intervals of $\delta u=0.2$. \label{fig:charfun_IBM}}
\end{figure}

It has been discussed in Section~\ref{sec:charfun} that in order to evaluate the statistical moments of the heat exchange, one needs to evaluate the derivative of the characteristic function at point $u=0$. Some symmetries of $\chi(u)$ prove to be useful in this numerical calculation. Specifically, from the definition in Eq.~\eqref{eq:char_func}, it is clear that
\begin{equation}
    \label{chi_symmetry}
   \chi^{*}(u)=\chi(-u).
\end{equation}
since the probability distribution $P(Q)$ is a real function. This implies that the real and imaginary parts of $\chi(u)$ have the symmetries
\begin{align}
    Re(\chi(u)) & = Re(\chi(-u))\nonumber, \\
    Im(\chi(u)) & = -Im(\chi(-u)).\label{real_imag_prop}
\end{align}
and are shown in Fig.~\ref{fig:charfun_IBM}, for both the independent-boson and spin-boson model.
Note that in Fig.~\ref{fig:charfun_IBM} it was not computationally possible to evaluate the characteristic function up to equilibrium time for values of $u$ higher than those represented. In the case of the spin-boson model, $t=9.53$ was the maximum time the TEMPO algorithm was able to reach for $u=3$.\\
In our method we perform a numerical differentiation in order to calculate the first and second moments of the heat distribution, as shown in Eq.~\eqref{eq:heat_moments_chi}. In order to do that, we have to choose a suitable value of $u$. Note however that the counting field is not a numerical parameter of the TEMPO algorithm, but a variable of the characteristic function. For the mean heat
\begin{equation}
\langle Q \rangle = -i\left.\frac{dRe\left[\chi\left(u\right)\right]}{du}\right|_{u=0}+\left.\frac{dIm\left[\chi\left(u\right)\right]}{du}\right|_{u=0}\label{mean_numerical},
\end{equation}
and it is clear from Fig.~\ref{fig:charfun_IBM} that $dRe\left[\chi\left(u\right)\right]/du|_{u=0}=0$, which can also be deduced from the symmetry properties in Eq.~\eqref{real_imag_prop}.\\
Since $\chi(0)=1$, then $Im[\chi(0)]=0$ and the numerical derivative in the right-hand side of Eq.~\eqref{mean_numerical}, evaluated for a small enough value $u_{\epsilon}$, is
\begin{equation}
\langle Q \rangle = \frac{Im[\chi(u_{\epsilon})]}{u_{\epsilon}}+\mathcal{O}\left(u_{\epsilon}\right)
\end{equation}
We find that the mean heat depends only on the imaginary part of the characteristic function and is given by the linear slope of the function depicted in Fig.~\ref{fig:charfun_IBM} in an interval $[0,u_{\epsilon}]$, with an error of the order $\mathcal{O}\left(u_{\epsilon}\right)$.  The value $u_{\epsilon}$ must be such that within the interval it defines, the real part of the characteristic function can still be approximated by a constant function, and the slope of the imaginary part is linear. $u_{\epsilon}$ will depend on the model, as shown by comparing the two figures in Fig.~\ref{fig:charfun_IBM}, and on the physical parameters $\alpha$, $T$ and $\omega_{C}$. Indeed, Fig.~\ref{fig:secondcumulantIBM} shows for example that while for $\alpha=0.1$ it is sufficient to take $u=0.01$, for stronger coupling such as $\alpha=1.5$ it is necessary to set $u=0.005$ to achieve the same precision.\\
We have found that in order to achieve a function $\langle Q \rangle$ that is constant in the long time limit, for the parameters considered in this work the value of  $u_{\epsilon}$ can't be greater than $u_{\epsilon}=0.01$. In general, decreasing the value of $u_{\epsilon}$ below $u_{\epsilon}=0.005$  will increase the computational time but not improve significantly the precision of the result.

\subsection{TEMPO memory depth}
In Section~\ref{sec:TEMPOalgorithm} we have discussed the finite memory depth $K$ of the TEMPO algorithm that allows it to efficiently propagate the ADT. In this subsection of the Appendix we will show how the memory depth affects the convergence of the mean heat and the variance of the heat distribution studied throughout this work.\\
The form of the correlation functions in Eq.~\eqref{eq:cont_etaC}-\eqref{eq:cont_etaA2} sets the minimum value of $K\Delta$ needed. Indeed, $K\Delta$ has to be large enough to so that the discretised correlation functions are zero. Preliminary calculations have shown that, for the values of temperature and coupling strength considered, this requirement is satisfied around the value $K\Delta=5$.
\begin{figure}
    \centering
    \includegraphics[scale=0.5]{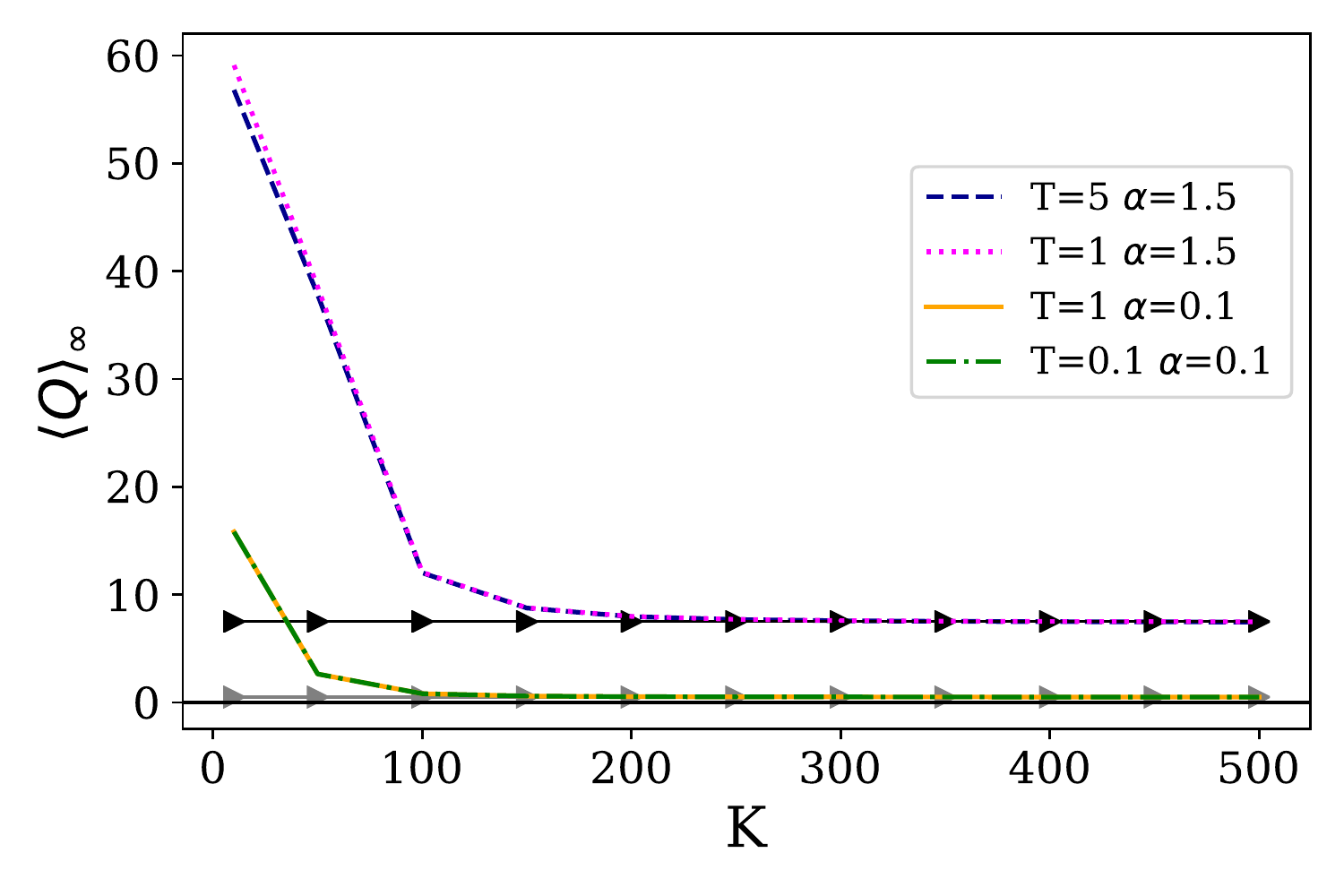}
    \includegraphics[scale=0.5]{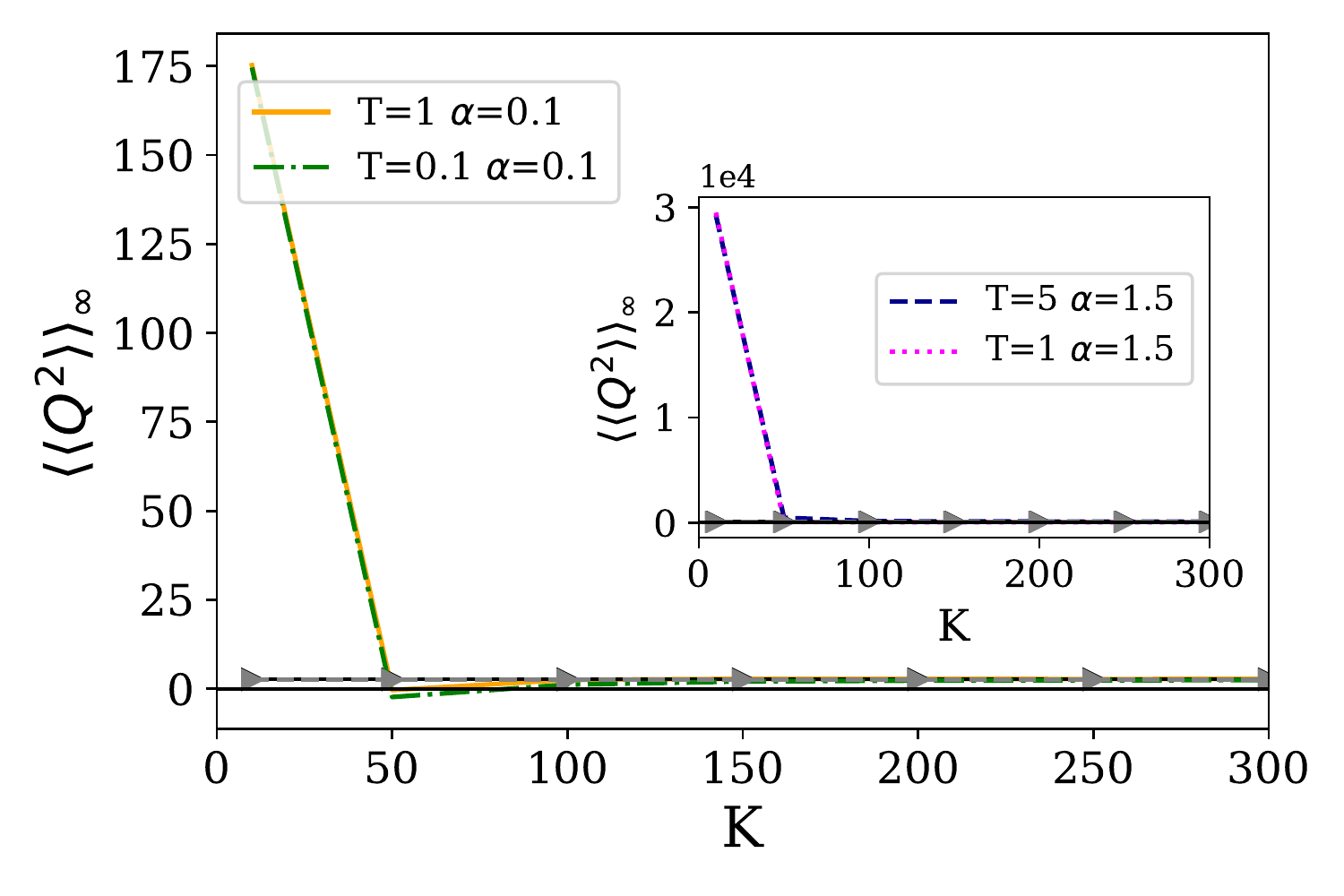}
    \caption{Upper figure: asymptotic mean heat for the independent-boson model as a function of $K$. Lower figure: asymptotic variance of the heat distribution for the independent-boson model as a function of $K$. Triangles represent the analytical solution given by Eq.~\eqref{mean_heat_IBM} (upper figure) and Eq.~\eqref{variance_IBM} (lower figure) in the long time limit. The figures are plotted for different values of the temperature and coupling strength. The remaining parameters are set to $\omega_{C}=5$, $\Delta=0.01$, $p=100$ and $u=0.01$.\label{fig:IBM_convergence}}
\end{figure}
Fig.~\ref{fig:IBM_convergence} shows that in the independent-boson model, for a fixed value of $\Delta$, both the mean heat and the variance of the heat distribution reach the predicted asymptotic value for $K>300$, for all the values of $T$ and $\alpha$ depicted. For values $K<100$, however, the asymptotic TEMPO result diverges greatly from the predicted one. This clearly shows how our method, which is able to operate at high values of the memory depth, has a much greater accuracy than other methods which operate in the region $K<100$.

\section{Variational theory of heat transfer}
\label{app:Silbey_Harris}

In this Appendix, we give details of the variational approach to describing heat transfer at low temperature~\cite{Silbey1984}. Applying the transformation in Eq.~\eqref{polaron_transformation}, we arrive at $\hat{P}^\dagger \hat{H}\hat{P} = \hat{H}_0' + \hat{H}_I'$, where 
\begin{align}
  \hat{H}'_0 & = \Omega'\hat{S}_x + \sum_j\omega_j \hat{a}_j^\dagger \hat{a}_j + \sum_j \frac{f_j(f_j-2g_j)}{4\omega_j}, \\
\label{Hint_variational}
    \hat{H}_I' & = \Omega \left[ \left(\hat{B}-B \right) \hat{S}_+ + {\rm h.c.}\right] + \hat{S}_z \sum_j (g_j-f_j)(\hat{a}_j + \hat{a}_j^\dagger).
\end{align}
Here, we defined the renormalised tunnelling amplitude $\Omega' = \Omega B$, where $B = \langle \hat{B}\rangle_{\hat{H}_0'} \equiv \Tr[\hat{B} e^{-\beta \hat{H}_0'}]/Z_0'$, with $Z_0' = \Tr[e^{-\beta \hat{H}_0'}]$ and
\begin{align}
    \label{B_def}
    \hat{B} & = \prod_j \exp\left[ \frac{f_j}{\omega_j}\left(\hat{a}_j^\dagger -\hat{a}_j\right)\right],
\end{align}
while $\hat{S}_+ = (\hat{S}_x + i \hat{S}_y)/2$ is the spin raising operator. Carrying out the thermal average explicitly, we find 
\begin{equation}
    \label{B_explicit}
    B = \exp\left[-\frac{1}{2}\sum_j\frac{f_j^2}{\omega_j^2}\coth\left(\frac{\beta\omega_j}{2}\right)  \right].
\end{equation}

The variational parameters $\{f_j\}$ are determined by minimising the Feynman-Bogoliubov upper bound on the free energy, $F = -T\ln \Tr[e^{-\beta\hat{H}}]$, given by
\begin{equation}
    \label{FB_bound}
    F \leq F_B = -T\ln Z_0' + \langle \hat{H}_I'\rangle_{\hat{H}'_0} + O(\langle \hat{H}_I'^2\rangle_{\hat{H}'_0}).
\end{equation}
Since $\langle \hat{H}_I'\rangle_{\hat{H}'_0}=0$ by construction, we find that
\begin{equation}
    \label{FB_explicit}
   F_B = \sum_j \frac{f_j(f_j-2g_j)}{4\omega_j} - T \ln \left[2 \cosh\left(\frac{\beta \Omega'}{2}\right) \right],
\end{equation}
where we have neglected higher-order terms in $\hat{H}_I'$ since this is small by assumption. The minimum defined by $\partial F_B/\partial f_j = 0$ is then easily found to be $f_j = g_j \phi(\omega_j)$, where $\phi(\omega)$ is given by Eq.~\eqref{phi_def}. Plugging this result into Eq.~\eqref{B_explicit} yields the renormalised tunnelling matrix element in Eq.~\eqref{B_solution}.\\
Using these results in Eq.~\eqref{Hint_variational} also shows self-consistently that $\hat{H}_I' = O(\Omega)$. Eq.~\eqref{FB_bound} can thus be interpreted as a formal expansion in powers of $\Omega/\omega_c$. That is, the variational approach treats the spin Hamiltonian $\hat{H}_S$ as a small perturbation with respect to the independent-boson Hamiltonian $\hat{H}_B + \hat{H}_I$, and becomes exact in the limit $\Omega\to 0$.

\clearpage

\end{document}